\documentclass[table,twocolumn,trackchanges]{aastex631} 
\usepackage{booktabs}
\usepackage{amsmath,bm}
\usepackage{multirow}

\submitjournal{AAS Journals}

\begin{document}

\title{A New Stellar Mass Proxy for Subhalo Abundance Matching}

\author[0000-0003-2069-9413]{Chen-Yu Chuang}
\affiliation{Institute of Astronomy, National Tsing Hua University, Hsinchu 30013, Taiwan}
\affiliation{Institute of Astronomy and Astrophysics, Academia Sinica,   Taipei 10617, Taiwan}

\author[0000-0001-7146-4687]{Yen-Ting Lin}
\affiliation{Institute of Astronomy and Astrophysics, Academia Sinica,   
Taipei 10617, Taiwan}

\begin{abstract}

Subhalo abundance matching (SHAM) has played an important role in improving our understanding of how galaxies populate their host dark matter halos.  In essence, the SHAM framework is to find a dark matter halo property  that best correlates with an attribute of galaxies, such as stellar mass.  The peak value of the maximum circular velocity ($V_{\rm max}$) a halo/subhalo has ever attained throughout its lifetime, $V_{\rm peak}$, has been a popular choice for SHAM. A recent study by \citet{Tonnesen2021} suggested that quantity $\phi$, which combines the present-day $V_{\rm max}$ and the peak value of halo dark matter mass, performs better in predicting stellar mass than $V_{\rm peak}$. Inspired by their approach, in this work, we find that further improvement can be achieved by a quantity $\psi_5$ that combines the 90th percentile of $V_{\rm max}$ a halo/subhalo has ever achieved with the 60th percentile of the dark matter halo time variation rate. Tests based on the simulation {\it TNG300 of the IllustrisTNG project} show that our new SHAM scheme, with just three free parameters, can improve the stellar mass prediction and mass-dependent clustering by 15\% and 18\% from $\phi$, respectively, over the redshift range $z=0-2$.
\end{abstract}


\section{Introduction} \label{sec:intro}

Among the methods employed to study the galaxy--halo connection (in simplest terms, how galaxies populate dark matter halos; see \citealt{wechsler18} for a recent review), subhalo abundance matching (SHAM; see e.g., \citealt{kravtsov04,tasitsiomi04,vale04,conroy06,Stiskalek2021}; to name only a few) is popular for its simplicity -- typically only involving a couple of free parameters, the model galaxies produced by the method by construction match the observed galaxy stellar mass (or luminosity) function, 
as well as the stellar (or luminosity) dependence of the two-point correlation function (2PCF).

In the SHAM framework, one essentially is seeking a dark matter halo property $X$ that is most tightly correlated with a galaxy property (typically the stellar mass, $M_\star$).  It has been a popular choice to set $X=V_{\rm peak}$,  the peak value of the maximum circular velocity ($V_{\rm max}$) a halo (or subhalo) has ever attained. With the inclusion of some scatter $\sigma_V$ between $V_{\rm peak}$ and $M_\star$, the model is shown to be able to best reproduce the observed 2PCFs of galaxies selected above different stellar mass thresholds, compared to models that use other properties as $X$, such as peak (sub)halo mass throughout its lifetime ($M_{\rm DM, peak}$ e.g., \citealt{reddick13,zentner14}; but see \citealt{masaki22}).

\citet{lehmann17} were among the first to consider $X$ to be a combination of multiple halo properties.
Recently, \citet[][hereafter TO21]{Tonnesen2021}  proposed a new property, $X=\phi\equiv V_{\rm max,0}/v_{\rm max,0@12.7}+M_{\rm DM, peak}/(10^{12.7}M_\odot$) (please refer to equation \ref{eq:5} below for the definition of these terms), 
which is shown to be an excellent stellar mass proxy.
Inspired by their approach, here we explore variants to $\phi$, and propose a new property $\psi_5$ that further improves the accuracy in stellar mass prediction by 15\% with respect to that of $\phi$.

This paper is structured as follows. In Section~\ref{sec:data} we describe the simulation and the loss metrics used for our abundance matching (AM) scheme. Then in Section~\ref{sec:rank} we present our methodology for finding the best stellar mass indicator, which is referred to as $\psi_5$.  Our results are presented in Section~\ref{sec:res}, where we show the performance of $\psi_5$, in terms of reproducing the stellar mass and color dependencies of galaxy clustering for mock galaxies from {\it TNG300 of the IllustrisTNG project} \citep[][hereafter TNG300]{Marinacci2018,Naiman2018,Nelson18,pillepich18,springel18,Nelson2019} from $z=0-2$.  We discuss and summarize our results in Section~\ref{sec:disc} and Section~\ref{sec:con}, respectively.

\section{Simulation and Methods}
\label{sec:data}
We test alternatives to $\phi$ using the TNG300 simulation suites, and quantify the performance by the prediction of $M_\star$ and stellar mass/color-dependent 2PCF. In this section, we first describe how we select galaxies from the simulation, then explain how  the 2PCF is measured, 
as well as how the performance of the variants of $\phi$ (which shall be referred to as $\psi$) is evaluated.

\subsection{TNG300}
\label{sec:sim}
We extract the dark matter and baryonic features from TNG300, which has a box of $300\,\rm{Mpc}$ on a side and contains $(2500)^3$ dark matter particles and gas particles. To select the well-resolved galaxies, we choose model galaxies with stellar mass\footnote{From the SubhaloMassType field in the TNG300 catalog.} $M_\star\geq10^9\,h^{-1}M_\odot$ and the dark matter subhalos with dark matter mass $M_{DM}\geq10^{11}\,h^{-1}M_\odot$ in the dark matter only (DMO) counterpart of TNG300, where h=0.6774. Then, we separate the galaxies into satellites and centrals using the subhalo catalog of TNG300.

Following TO21, we abundance match the stellar mass for the \emph{all}, \emph{central}, and \emph{satellite} groups of galaxies, while a combined sample in which \emph{central}, \emph{satellite} are fit separately is labeled as the \emph{mixed} group. To reproduce the color dependence of the 2PCFs (see Section \ref{sec:clust}, we also separate the subhalos into blue and red samples by the color of the galaxies they host:
\begin{equation} \label{eq:col}
(u-g)_{\rm cut}=-0.031M_g-0.065z+0.695
\end{equation}
where the $u-g$ is the Sloan Digital Sky Survey (SDSS) $u-g$ color, $M_g$ is the $g$-band absolute magnitude, and $z$ is the redshift \citep[following][]{Skibba14}. We obtain the magnitudes of subhalos from the catalog published by \citet[][]{Nelson18}.

\subsection{2PCF}
\label{sec:clust}

A 2PCF, $\xi(r)$, is a measure of excess probability of finding a pair of galaxies separated by a distance $r$ over a random distribution. Observationally, to avoid contamination from peculiar velocities along the line of sight (LOS), one typically integrates along the redshift dimension and obtains the projected 2PCF, defined as
\begin{equation} \label{eq:1}
w_p(r_p)=\int^\infty_{-\infty}\xi(r_p,\pi)d\pi\approx  2\times \int^{\pi_{max}}_0\xi(r_p,\pi)d\pi
\end{equation}
where $r_p$ is the separation on the plane of the sky and $\pi$ is the  distance along the LOS.
The upper limit of LOS integration, $\pi_{max}$, is chosen to be $40\,$Mpc, which is consistent with the choice of actual measurements done with real data \citep[e.g.,][]{lin16,lin22}.

In Sections~\ref{sec:resclust} \&  \ref{sec:resclust_col}, we compare the 2PCFs from the true galaxy position and that from the abundance-matched galaxy position in different $M_\star$ and color bins. We make use of a high-performance package, {\tt Corrfunc} \citep[][]{Sinha2020, Sinha_inprep}, for the $w_p$ calculations.

\subsection{Loss Measures}
\label{sec:galsamp}

The primary objective of our study is to predict the $M_\star$ of each subhalo accurately while reproducing $w_p$ in the simulation. To measure the difference between the prediction and the ground truth, we define three  metrics. We use the scatter to gauge the $M_\star$ predictions:
\begin{equation} \label{eq:2}
\sigma=\sqrt{\frac{1}{N}\sum^N_i(\Delta y_i-\overline{\Delta y})^2}
\end{equation}
where $N$ is the total number of galaxies, $\Delta y_i=
\log(M_{\star,i})-\log\widehat{M_{\star,i}}$ is the residual of single $M_\star$ prediction and $\overline{\Delta y}$ is the average of the residual. To compare our results with TO21, we also calculate the error, defined as
\begin{equation} \label{eq:3}
{\rm{Error}}=\frac{\sum^N_i|\log(M_{\star, i}/\widehat{M_{\star,i}})|}{N}
\end{equation}

For $w_p$ predictions, we obtain the average of the fractional error for the $r_p$  and $M_\star$ bins. 
\begin{equation} \label{eq:4}
L=\frac{1}{N_{M_\star}\times N_{r_p}}\sum_{N_{M_\star}}\sum_{N_{r_p}}|\frac{w_p(r_p, M_\star)-\widehat{w_p}(r_p, M_\star)}{w_p(r_p, M_\star)}|
\end{equation}
where $N_{M_\star}$ and $N_{r_p}$ are the number of $M_\star$ and $r_p$ bins, respectively. The uncertainty is measured by jackknife resampling. We calculate the $w_p$ with each jackknife sample and obtain the uncertainty via 
\begin{equation} \label{eq:werr}
\sigma_{w}(w_p) = \sqrt{\frac{N-1}{N}\sum^N_j(w_p-\widehat{w_{p,j}})^2}
\end{equation}
where $N=27$ is the number of jackknife samples, and $\widehat{w_{p,j}}$ is the projected 2PCF associated with the given jackknife sample. Here we do not include scatter in our $w_p$ estimation.

\section{Rank Ordering Schemes}
\label{sec:rank}

The simplest SHAM scheme is assuming a one-to-one correspondence between galaxies and dark matter (sub)halos of the same number density, where the rank order is through $M_\star$ for galaxies, and dark matter mass for (sub)halos. However, the dark matter mass of subhalos usually suffers from tidal stripping, and thus is not able to adequately reflect the rank order of stellar mass. TO21 improved this by introducing two parameters, the $V_{\rm max}$ at $z=0$ ($V_{\rm max,0}$) and $M_{\rm DM, peak}$. These properties are shown to be less sensitive to tidal stripping than dark matter mass. By combining these parameters into a single dimensionless parameter 
\begin{equation} \label{eq:5}
\phi=V_{\rm max,0}/V_{\rm max,0@12.95}+M_{\rm DM, peak}/10^{12.95}\, \rm{M_\odot} 
\end{equation}
(a \emph{pivot mass} of $M_{\rm DM, peak}=10^{12.95}\, M_\odot$ is introduced to normalize $\phi$; $V_{\rm max,0@12.95}$ is $V_{\rm max,0}$ for a halo of the pivot mass.\footnote{The mass is the sum of gravitationally bounded particles as provided by the {\tt SUBFIND} algorithm \citep{Springel2001}}), TO21 achieved an improvement of 27\% compared to using dark matter mass alone for the $M_\star$ prediction.\footnote{As TO21 worked on TNG100 of the IllustrisTNG project (with a box size of $100^3$\,Mpc$^3$), we have to refit the pivot mass.  Using their original pivot mass ($M_{\rm DM, peak}=10^{12.7}\, M_\odot$), the improvement is 34\% (TO21).} In our work, we try to compare $\phi$ with the $X$ we developed with TNG300. Thus, we refit the pivot mass for $\phi$ in the simulation and find $M_{\rm DM,peak}=10^{12.95}\,M_\odot$ for redshift 0.

Our goal is to seek other parameter combinations $\psi$ that perform better than $\phi$, particularly for the clustering properties, an aspect not addressed fully in TO21. To ensure a fair comparison between $\psi$ and $\phi$, we do not include scatters between $M_\star$ and the dark matter subhalo properties in our main results, as  TO21 did not consider such a scatter. However, we do present results with scatter in Appendix \ref{sec:appa}, showing that our best $\psi$ does not change when scatters are introduced.
Although the maxima of subhalo properties throughout their lifetimes such as $V_{\rm peak}$ and $M_{\rm DM, peak}$ are shown to better trace $M_\star$, the stochastic nature of those features as seen from the simulations can cause additional scatter. Thus, instead of using the maximum (i.e., 100th percentile) of those properties, we use a specific percentile for those properties to avoid the scatter due to stochasticity. We tried the 10th, 20th, ... 100th ($=$\,peak) percentiles for  $V_{\rm max}$ and $M_{\rm DM}$ throughout the lifetimes of subhalos, and conducted abundance matching using different percentiles of each property. It is found that the 90th percentile can predict $M_\star$ with the smallest scatter among all cases considered for $V_{\rm max}$ and $M_{\rm DM}$ (hereafter $V_{\rm max,90\%}$ and $M_{\rm DM,90\%}$).\footnote{With such a notation, $V_{\rm peak} \equiv V_{\rm max,100\%}$.} Replacing the properties in $\phi$ with the percentile properties, we found that the scatter of abundance matching decreases by $\sim12.5\%$.

Motivated by the finding that the peak value of the quantities often used in SHAM does not always lead to the best results, we then conducted a thorough examination of the performance of {\it all} the dark matter subhalo properties provided by TNG300 of different percentiles. The best alternatives of $V_{\rm max,90\%}$ is $V_{\rm disp,80\%}$ while those of $M_{\rm DM,90\%}$ are $M_{R_{\rm max},\rm 80\%}$ and $|\dot{M}_{\rm DM}|_{\rm 60\%}$, where $V_{\rm disp}$ is the velocity dispersion of a subhalo, $M_{R_{\rm max}}$ is the dark matter mass within the radius that corresponds to $V_{\rm max}$, and $|\dot{M}_{\rm DM}|$ is the absolute subhalo dark matter mass variation rate. We will show in Section \ref{sec:res} that 
$X$ based on the combinations of these quantities can improve the prediction of stellar mass and $w_p$, and  define six  candidates for $X$ as
\begin{equation} \label{eq:6}
\psi_1=\frac{V_{\rm max,90\%}}{V_{\rm max,90\%@13.3}}+\frac{M_{\rm DM,90\%}}{M_{\rm DM,90\%@13.3}}
\end{equation}

\begin{equation} \label{eq:7}
\psi_2=\frac{V_{\rm disp,80\%}}{V_{\rm disp,80\%@13.6}}+\frac{M_{\rm DM,90\%}}{M_{\rm DM,90\%@13.6}}
\end{equation}

\begin{equation} \label{eq:8}
\psi_3=\frac{V_{\rm max,90\%}}{V_{\rm max,90\%@13.5}}+\frac{M_{R_{\rm max},\rm 80\%}}{M_{R_{\rm max},\rm 80\%@13.5}}
\end{equation}

\begin{equation} \label{eq:9}
\psi_4=\frac{V_{\rm disp,80\%}}{V_{\rm disp,80\%@13.8}}+\frac{M_{R_{\rm max},\rm 80\%}}{M_{R_{\rm max},\rm 80\%@13.8}}
\end{equation}

\begin{equation} \label{eq:10}
\psi_5=\frac{V_{\rm max,90\%}}{V_{\rm max,90\%@13.2}}+\frac{|\dot{M}_{\rm DM}|_{\rm 60\%}}{|\dot{M}_{\rm DM}|_{\rm 60\%@13.2}}
\end{equation}

\begin{equation} \label{eq:11}
\psi_6=\frac{V_{\rm disp,80\%}}{V_{\rm disp,80\%@13.7}}+\frac{|\dot{M}_{\rm DM}|_{\rm 60\%}}{|\dot{M}_{\rm DM}|_{\rm 60\%@13.7}}
\end{equation}
where the parameters with a subscript $\rm @$ are the normalization factors at a fitted pivot $M_{\rm DM,peak}$ right after. We tried a range of pivot masses ($M_{\rm DM, peak}=10^{10.0}-10^{14.0}\, M_\odot$) and select the one that minimizes the scatter of $M_\star$ prediction for each $\psi$'s. The subhalo properties along the main primary branch are used to calculate $\dot{M}_{\rm DM}$ and all the parameters with the subscript of ``peak'' or a certain percentile.

\section{Results}
\label{sec:res}

\subsection{Stellar Mass Prediction}
\label{sec:strmas}

\begin{figure*}
  \centering
  \includegraphics[width=0.8\linewidth]{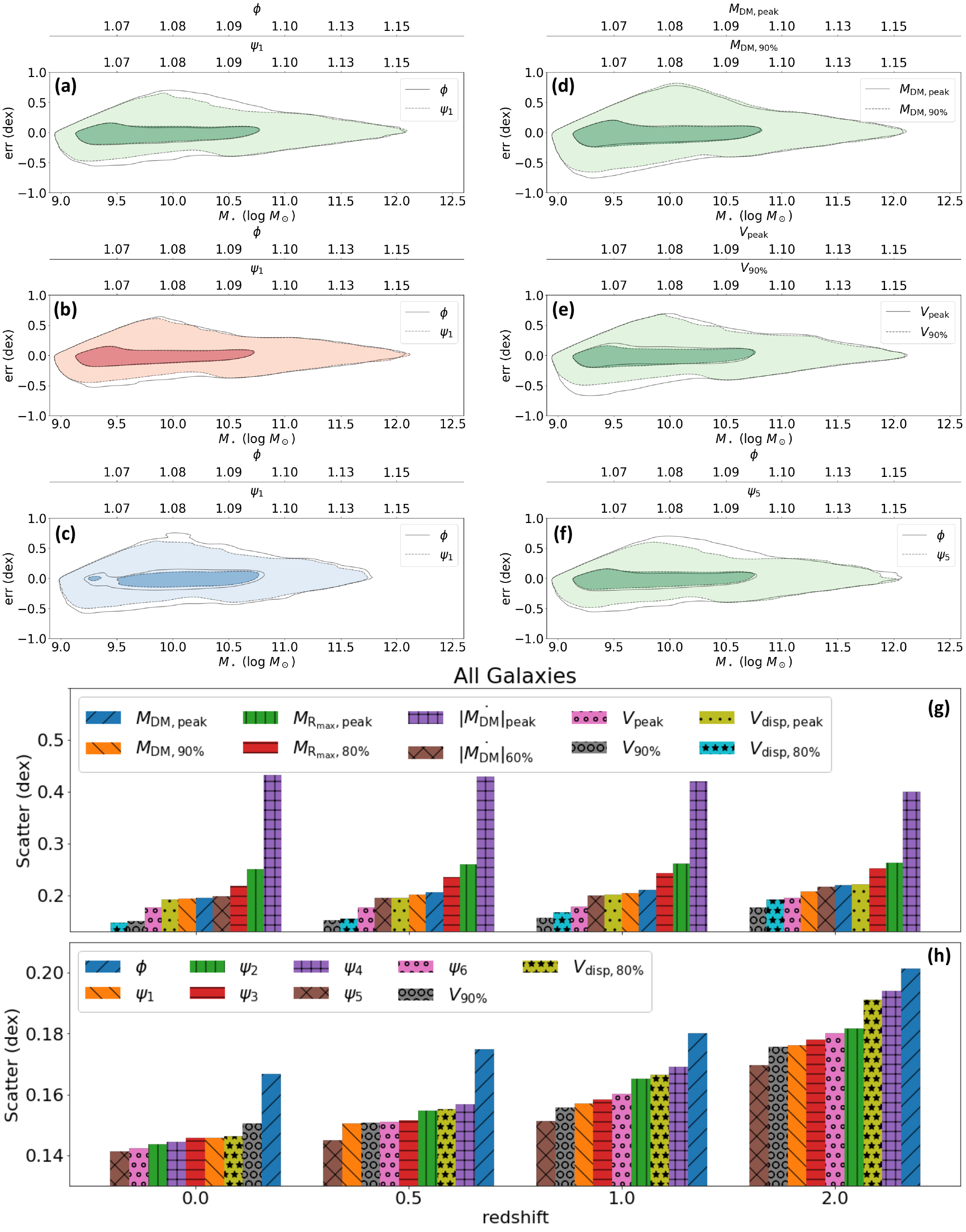}
  \caption{Panels (a) - (e): the error of the stellar mass prediction corresponding to stellar mass using the AM scheme of $\phi$ and $\psi_1$ for panels (a) - (c), $M_{\rm DM,peak}$ and $M_{\rm DM,90\%}$ for panel (d), $V_{\rm peak}$ and $V_{\rm max, 90\%}$ for panel (e), and $\phi$ and $\psi_5$ for panel (f).
  The green, red, and blue contours represent the predictions of ``all'', ``central'' and ``satellite'' groups, respectively. There are two contour lines for each AM scheme, which denote the data point density of 0.05 and 0.5 (normalized to maxima of 1). Panels (g) and (h): the overall scatter at several redshifts with different AM schemes denoted on the top left of the panel. The bars for each redshift bin are sorted so that the AM scheme with the lowest scatter is on the left.
  }
  \label{fig:scat}
\end{figure*}

We show in Table \ref{tab:comp}  the performance of each AM scheme in stellar mass prediction. In the upper-left part of each redshift of the table, the lifetime peak properties are improved by selecting an optimized percentile for all four groups of galaxies mentioned in Section \ref{sec:sim}. In the lower-left part of each redshift, all $\psi$'s show a relatively better $M_\star$ prediction than the original $\phi$ for the four groups. It is also shown that $\psi_1$ performs slightly worse than the other $\psi$'s, which means a better prediction of $M_\star$ can be achieved by replacing $V_{\rm max,90\%}$ and $M_{\rm DM, 90\%}$ with different properties. 

In Figure~\ref{fig:scat}, we compare the prediction errors with respect to true $M_\star$ using different AM schemes. The contours show the density distribution of the data points in the panels. A well-performed AM scheme should have contours closer to the horizontal line at zero error. Panel (a) of the Figure shows that the main improvement from $\phi$ to $\psi_1$ is for $\log (M_\star/M_\odot)<11.0$.  Panels (b) and  (c) show that the improvement in central galaxies is limited to lower-mass galaxies, while that of the satellites also occurs in higher-mass galaxies as well. Panels (d) and (e)  show the AM performances of individual parameters. Here, we can see that the improvement from $M_{\rm peak}$ to $M_{\rm DM, 90\%}$ is lower than that of $V_{\rm peak}$ to $V_{\rm max, 90\%}$. Panel (f)  shows the error distribution of $\psi_5$, which has the best performance among all AM schemes considered here. Finally, panels (g) and (h) show the scatter of all AM schemes at several different redshifts. We can see that the scatters of all the $\psi$'s are smaller than that of $\phi$ at all redshifts.

\subsection{Stellar Mass Dependence of Projected 2PCF}
\label{sec:resclust}

\begin{figure*}
  \centering
  \includegraphics[width=0.75\linewidth]{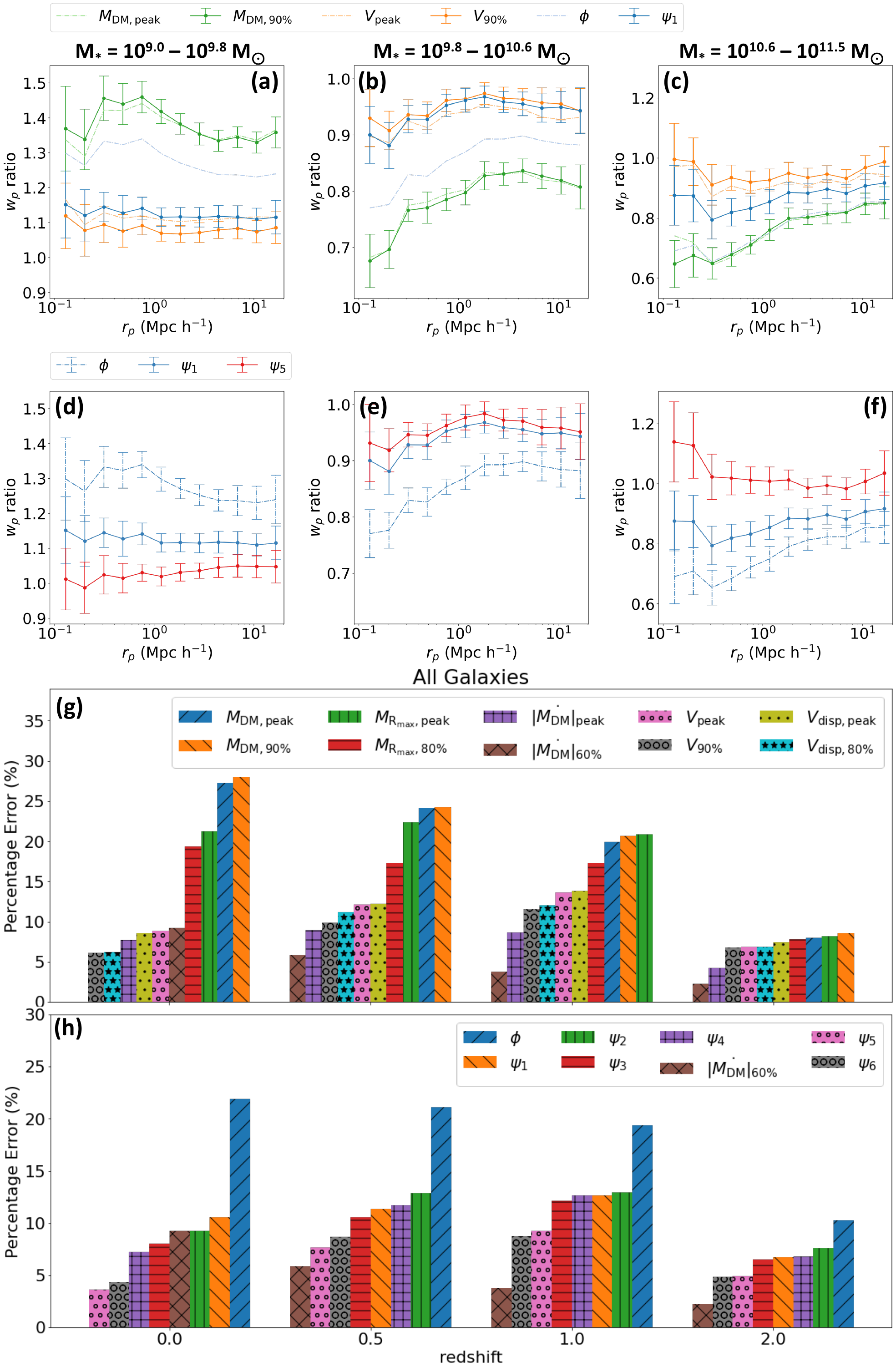}
  \caption{Panels (a) to (f): The data points denote the rational difference in the projected 2PCF between the true and abundance-matched galaxy distribution while the error bars are the jackknife uncertainties. The mass range of each stellar mass bin is denoted at the top of each panel. (g) and (f) show the fractional difference in the clustering averaged over all $r_p$ bins and $M_\star$ bins in the ``all'' group. The bar at each redshift is sorted the same way as Figure\ref{fig:scat}.
  The data points with jackknife uncertainties larger than 0.2 are omitted in all panels.}
  \label{fig:clust_col}
\end{figure*}

In Table \ref{tab:comp}, the upper-right part in each redshift of the table shows that the peak properties have a larger error for projected 2PCF prediction than the percentile properties, except for the $M_{\rm DM}$ properties and the $|\dot{M}_{\rm DM}|$ properties at $z=0$. In the lower-right of each redshift, we can also find that the $\psi$'s predictions of $w_p$ are better than that of $\phi$. Figure \ref{fig:clust_col} shows the ratio of the predicted $w_p$ over the true $w_p$. The first row (panels (a) - (c)) and the second row (panels (d) - (f))  compare the performances between the peak valued quantities to the percentile quantities and that between $\phi$ to $\psi$'s, respectively. We see that the differences in $w_p$ ratios are similar among all the $M_\star$ bins. Panels (g) and (h) show that while $\psi_5$ and $\psi_6$ yield  the best two $w_p$ predictions at $z=0$, the $|\dot{M}_{\rm DM}|_{\rm 60\%}$ outperforms $\psi_5$ and $\psi_6$ at all the other redshifts by a small margin.

\subsection{Stellar Mass and Color Dependence of Projected 2PCF}
\label{sec:resclust_col}
The upper-right part in each redshift of Table \ref{tab:comp} shows the error of $w_p$ predictions for the red and blue galaxies. For $M_{\rm DM}$, replacing the peak values with percentile quantities worsens the $w_p$ predictions for all galaxies. However, for the other features, this replacement worsens the $w_p$ predictions for the red galaxies at multiple redshifts, while improving the predictions for the blue galaxies across all epochs. In the bottom right in each redshift, the $\psi$'s outperforms $\phi$ in both types of galaxies.

\begin{figure*}
  \centering
  \includegraphics[width=0.8\linewidth]{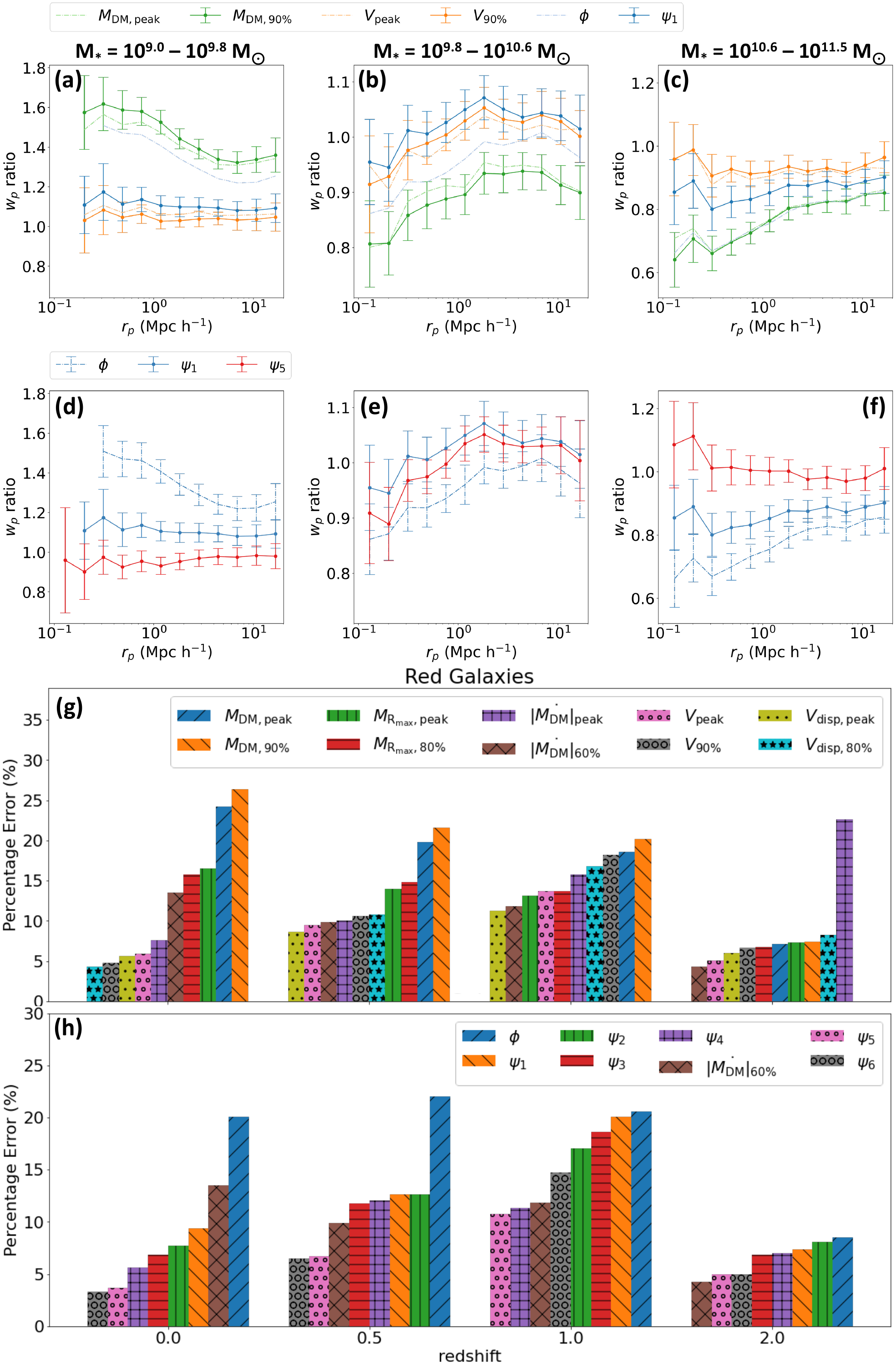}
  \caption{Same as Figure \ref{fig:clust_col}, but for red galaxies.
}
  \label{fig:clust_red}
\end{figure*}

\begin{figure*}
  \centering
  \includegraphics[width=0.75\linewidth]{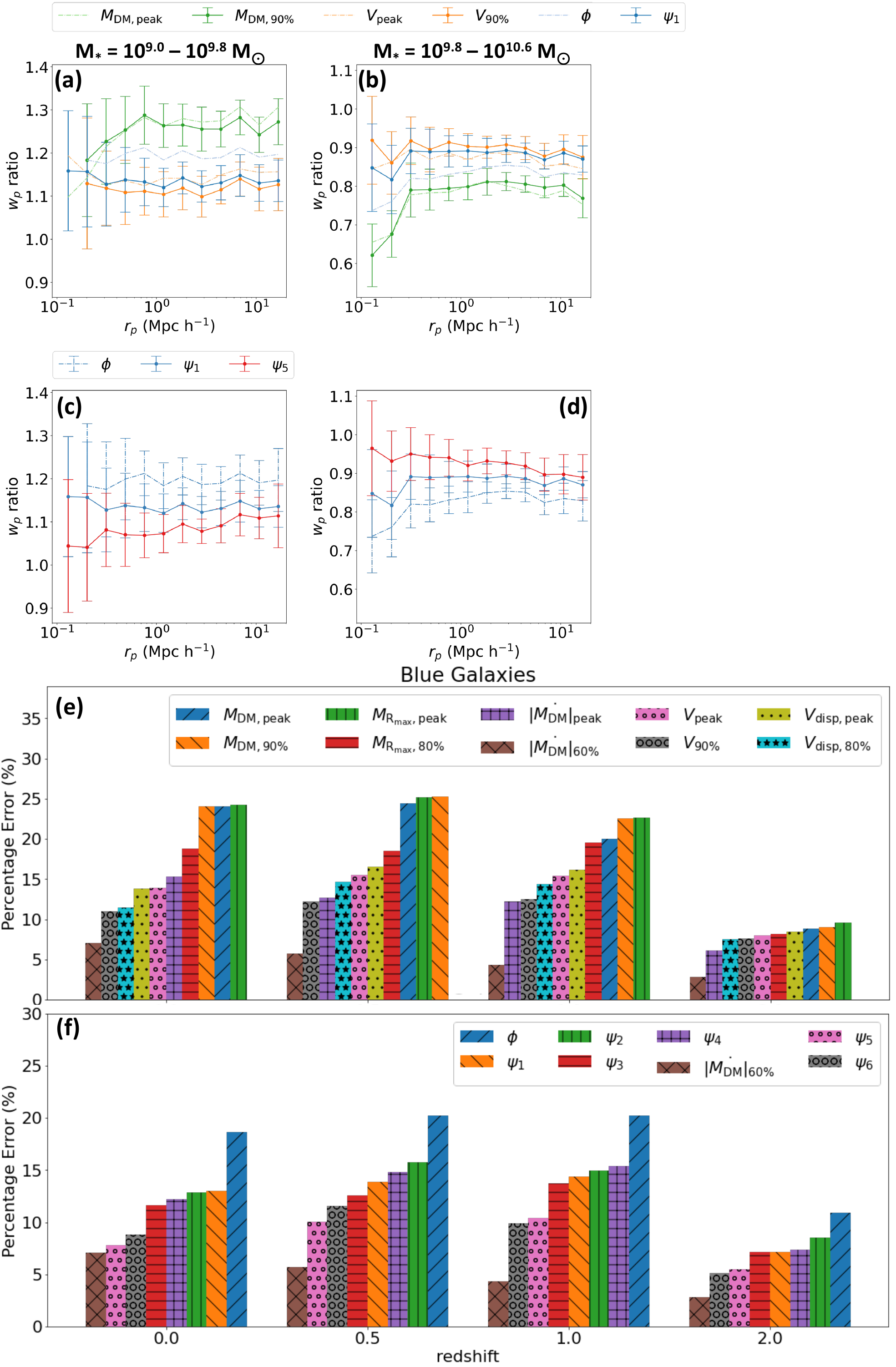}
  \caption{Same as Figure \ref{fig:clust_col}, but for blue galaxies. There are not enough galaxy pairs in the highest $M_\star$ bin at $z=0$ for reliable error estimation, forcing us to skip the bin.}
  \label{fig:clust_blue}
\end{figure*}

Figure \ref{fig:clust_red} and \ref{fig:clust_blue} show the same quantities as Figure \ref{fig:clust_col}, but for red and blue galaxies, respectively. The same as in Figure \ref{fig:clust_col}, the differences in $w_p$ ratios are similar among all the $M_\star$ bins. For the $w_p$ prediction of red galaxies, $M_{\rm DM,peak}$ performs better than $M_{\rm DM,90\%}$ in all $M_\star$ bins while $V_{\rm peak}$ performs roughly the same as $V_{\rm max,90\%}$. In the middle $M_\star$ bin, $\psi_1$ performs roughly the same as $\psi_5$, while $\psi_5$ performs better in other $M_\star$ bins. For most of the redshifts, $\psi_5$ and $\psi_6$ are the best in predicting $w_p$ for red galaxies. For the $w_p$ prediction of blue galaxies, $M_{\rm DM,90\%}$ and $V_{\rm max,90\%}$ perform better than $V_{\rm peak}$ and $M_{\rm DM,peak}$ in the lower two $M_\star$ bins. The performance of all properties in the highest $M_\star$ bins is not shown due to the small number of galaxy pairs. For most of the redshifts, $|\dot{M}_{\rm DM}|_{\rm 60\%}$ is the best for $w_p$ prediction of blue galaxies, outperforming $\psi_5$ with a small margin. Overall, $\psi_5$ is still reliable for reproducing color-dependent clustering.

\section{Discussion}
\label{sec:disc}

Using the halos and mock galaxies from TNG300,
we have shown that our best stellar mass proxy, $\psi_5$, performs better compared to both the popular choice of $V_{\rm peak}$, as well as the parameter $\phi$ proposed by TO21, in terms of predicting the stellar mass.  Furthermore, when used in SHAM, $\psi_5$ can reliably reproduce both the stellar mass and color dependencies of the 2PCFs at $z=0-2$ better than $V_{\rm peak}$. Whereas the usual implementation of $V_{\rm peak}$ model has one free parameter (the scatter between $V_{\rm peak}$ and $M_\star$; see below), our $\psi_5$ model has three (the percentiles for $V_{\rm max}$ and $|\dot{M}_{\rm DM}|$, and the normalizing halo mass). Nonetheless, the better performances in $M_\star$ prediction (and hence the stellar mass dependence of $w_p$) should outweigh the extra degrees of freedom.

In Section \ref{sec:rank}, we argued that the success of our method is attributed to the use of percentile quantities, which effectively smooths  the stochastic behavior of dark matter subhalo properties. Here, we provide a more quantitative analysis of the evolution of $M_{\rm DM},\ M_{\rm R_{max}},\ |\dot{M}_{\rm DM}|,\ V_{\rm max}$ and $V_{\rm disp}$ for the subhalos. For each quantity mentioned above, we first obtain the moving average with a window size of 10 snapshots along the time axis for each subhalo, and subtract the original values with averaged values to obtain the residual of property history. Next, we perform a fast Fourier transform to the residual to obtain an amplitude and frequency relation. Lastly, we calculate the average frequency for each subhalo weighted by the amplitude. We define this value as the stochasticity of a subhalo for a certain dark matter property. Now, we can compare the average stochasticity ($S$, Gyr$^{-1}$) of the properties with the relative improvement from the chosen  percentile quantities instead of the peak values for each property. In Table~\ref{tab:comp}, the relative improvement in scatter in the \emph{All} group for each property can be obtained directly ($I_\sigma=(\sigma_{\rm peak}-\sigma_{\rm perc})/\sigma_{\rm peak}$, where $\sigma_{\rm peak}$ and $\sigma_{\rm perc}$ are the scatter with peak and percentile quantities, respectively):
\begin{enumerate}
    \item $|\dot{M}_{DM}|$: $S = 1.843,\ I_\sigma=0.542$
    \item $V_{\rm disp}$: $S=1.419,\ I_\sigma=0.236$
    \item $V_{\rm max}$: $S=1.509,\ I_\sigma=0.15$
    \item $M_{R_{max}}$: $S=1.487,\ I_\sigma=0.124$
    \item $M_{\rm DM}$: $S=1.284,\ I_\sigma=0.013$
\end{enumerate}
The stochasticities are overall positively correlated with the improvements of switching from peak to percentile quantities. We can see that the properties with maximal ($|\dot{M}_{DM}|$) and minimal ($M_{\rm DM}$) stochasticity correspond to the largest and least relative improvements, respectively. A higher time resolution of the simulation might affect the best-fit percentiles of various properties considered here, especially for those with high stochasticities. However, due to the limitation of the TNG300 time resolution, we are not able to further confirm this scenario.

One caveat for the success of the color dependence of clustering is that we have separated red subhalos from blue ones {\it a priori}, based on the full hydrodynamical run of TNG300, making our approach non-applicable to other DMO simulations.\footnote{Please note that our $\psi_5$ can still accurately predict  the stellar mass dependence of 2PCFs at $z=0-2$.}  In C.-Y.~Chuang et al.(~2022, in preparation) from private communication, a graph neural network \cite[GNN, e.g.][]{Jespersen22} model is used to emulate the TNG300 results, which enables one to \emph{port} the TNG hydrodynamical model to any \emph{N}-body simulations of similar resolutions as TNG300.  Of course, with the GNN model, one then does not really require SHAM anymore, as the GNN model provides accurate predictions for $M_\star$, $g-r$ color,  star formation rate, and several other properties, for galaxies with $M_\star \ge 10^9 M_\odot$ at $z=0-2$.

\section{Conclusion}
\label{sec:con}
In recent years, the conditional abundance matching (\citealt{hearin14}; see also \citealt{masaki13}) approach has been used to exploit secondary connections between galaxy and halo properties, which allows one to assign colors to model galaxies (e.g., \citealt{hearin17}).  However, such models may not be compatible with the observed color-dependence of the central galaxy stellar mass--halo mass relation \citep[SHMR; e.g.,][]{mandelbaum16}.
As our model is calibrated against TNG300, it can only reproduce the observed galaxy properties to the extent that the TNG300 model can. It is found that, while the SHMR (irrespective of galaxy color) in TNG300 agrees with the measurements of \citet{leauthaud12}, when central galaxies are split by color, only the red galaxy SHMR is reproduced.
Finally, we note that there may well be an intrinsic scatter for the $M_\star$ prediction through SHAM in the real universe, while the scatter associated with the \emph{best} SHAM scheme could be smaller than the intrinsic scatter. Thus, our $\psi_5$-based SHAM scheme  may inherit potential biases created by the numerical simulation we rely on.

Therefore, for a more realistic (and useful) application of the methodology developed in this work, in a future paper, we plan to tune our $\psi$'s by directly fitting to observations, such as the clustering measurements from SDSS and its subsequent surveys, as well as deeper surveys like DEEP2,  PRIMUS, VIPERS, and UltraVISTA (e.g., \citealt{coil08,zehavi11,marulli13,guo14,Skibba14,McCracken15,coi17}).

\begin{acknowledgments}
We thank the anonymous referee for the comments that improved the manuscript.  We also appreciate the helpful comments from Shogo Masaki, Yao-Yuan Mao, Christian Jespersen, and Hong Guo. We acknowledge support from the National Science and Technology Council of Taiwan under grant Nos. MOST 110-2112-M-001-004 and MOST 109-2112-M-001-005. Y.T.L. thanks I.H., L.Y.L., and A.L.L. for constant encouragement and inspiration. 

The numerical work was conducted on the high-performance computing facility at the Institute of Astronomy and Astrophysics in Academia Sinica (https://hpc.tiara.sinica.edu.tw). The IllustrisTNG simulations were undertaken with compute time awarded by the Gauss Centre for Supercomputing (GCS) under GCS Large-Scale Projects GCS-ILLU and GCS-DWAR on the GCS share of the supercomputer Hazel Hen at the High Performance Computing Center Stuttgart (HLRS), as well as on the machines of the Max Planck Computing and Data Facility (MPCDF) in Garching, Germany.

\end{acknowledgments}

\setlength{\tabcolsep}{3pt}

\movetabledown=7.5cm
\begin{rotatetable*}
\begin{deluxetable*}{||l||l|llllllll|lll||l|llll|lll}
\tabletypesize{\footnotesize}

\tablecaption{Scatter and loss of stellar mass prediction via different abundance matching schemes. The best values are bold-faced and color shaded.\label{tab:comp}}
\tablehead{
Loss Metrics& \multicolumn{1}{c}{}& \multicolumn{8}{c}{Stellar Mass Prediction} & \multicolumn{3}{c}{Prediction of $w_p$}& \multicolumn{1}{c}{}
& \multicolumn{4}{c}{Stellar Mass Prediction} & \multicolumn{3}{c}{Prediction of $w_p$}\\
\midrule
Galaxy Types&\multicolumn{1}{c}{}& \multicolumn{2}{c}{All}& \multicolumn{2}{c}{Central}& \multicolumn{2}{c}{Satellite}& \multicolumn{2}{c}{Mix}&All & Blue & \multicolumn{1}{c}{Red}&\multicolumn{1}{c}{}
& All& Central& Satellite& \multicolumn{1}{c}{Mix}& \multicolumn{1}{c}{All} & Blue & Red\\
\midrule
AM Scheme&\multicolumn{1}{c}{}& Scatter & Loss & Scatter & Loss & Scatter & Loss & Scatter & \multicolumn{1}{c}{Loss} & \multicolumn{3}{c}{Percentage Error}&\multicolumn{1}{c}{}
& \multicolumn{4}{c}{Scatter}& \multicolumn{3}{c}{Percentage Error}\\
}
\startdata
$M_{DM,peak}$& \multirow{17}{*}{\rotatebox[origin=c]{90}{redshift 0}}& 0.1952 & 0.1484 & 0.1755 & 0.1327 & 0.1942 & 0.1477 & 0.1800 & 0.1361& 27.3$\pm$4.1 & 24.0$\pm$6.2 & 24.2$\pm$5.6&\multirow{17}{*}{\rotatebox[origin=c]{90}{redshift 1.0}}&0.2053 & 0.1869 & 0.2082 & 0.1915 &24.2$\pm$4.3 & 24.4$\pm$6.9 & 19.8$\pm$9.0 \\
$M_{\rm DM,90\%}$ && 0.1926 & 0.1461 & 0.1691 & 0.1273 & 0.1917 & 0.1473 & 0.1745 & 0.1318 &28.0$\pm$4.0 & 24.0$\pm$5.3 & 26.4$\pm$5.6 &&0.2003 & 0.1795 & 0.2027 & 0.1845 &24.2$\pm$4.3 & 25.3$\pm$7.1 & 21.6$\pm$8.5\\
$M_{\rm R_{max},peak}$ && 0.2494 & 0.1908 & 0.2445 & 0.1871 & 0.2407 & 0.1838 & 0.2436 & 0.1864 &21.2$\pm$3.8 & 24.2$\pm$7.0 & 16.5$\pm$5.4 &&0.2601 & 0.2521 & 0.2545 & 0.2526 &22.4$\pm$4.3 & 25.2$\pm$6.6 & 14.0$\pm$8.7\\
$M_{\rm R_{max},80\%}$ && 0.2185 & 0.1660 & 0.2115 & 0.1604 & 0.2168 & 0.1653 & 0.2127 & 0.1615 &19.3$\pm$3.9 & 18.8$\pm$6.2 & 15.8$\pm$5.6 &&0.2344 & 0.2267 & 0.2358 & 0.2286 &17.3$\pm$4.3 & 18.5$\pm$7.0 & 14.9$\pm$8.1\\
$|\dot{M_{\rm DM}}|_{\rm peak}$ && 0.4318 & 0.3497 & 0.4385 & 0.3555 & 0.4060 & 0.3289 & 0.4313 & 0.3494 &7.7$\pm$3.8 & 15.3$\pm$7.6 & 7.6$\pm$5.8 &&0.4296 & 0.4328 & 0.4123 & 0.4287 &8.9$\pm$4.2 & 12.7$\pm$6.1 & 10.1$\pm$8.3\\
$|\dot{M_{\rm DM}}|_{\rm 60\%}$ && 0.1977 & 0.1516 & 0.1919 & 0.1464 & 0.1994 & 0.1532 & 0.1937 & 0.1480 &9.3$\pm$4.3 & \cellcolor{blue!25}\textbf{7.1}$\pm$5.9 & 13.5$\pm$5.4 &&0.1949 & 0.1926 & 0.1969 & 0.1935 &\cellcolor{blue!25}\textbf{5.8}$\pm$4.4 & \cellcolor{blue!25}\textbf{5.7}$\pm$6.0 & 9.9$\pm$8.4\\
$V_{\rm peak}$ && 0.1769 & 0.1363 & 0.1739 & 0.1338 & 0.1799 & 0.1394 & 0.1753 & 0.1351 &8.8$\pm$3.9 & 13.9$\pm$6.0 & 5.9$\pm$5.1 &&0.1765 & 0.1725 & 0.1767 & 0.1734 &12.1$\pm$4.1 & 15.5$\pm$6.3 & 9.5$\pm$8.7\\
$V_{\rm 90\%}$ && 0.1504 & 0.1149 & 0.1469 & 0.1120 & 0.1565 & 0.1200 & 0.1491 & 0.1138 &6.1$\pm$3.8 & 11.0$\pm$5.6 & 4.7$\pm$5.1 &&0.1508 & 0.1471 & 0.1524 & 0.1482 &9.9$\pm$4.5 & 12.2$\pm$7.0 & 10.6$\pm$8.5\\
$V_{\rm disp,peak}$ && 0.1914 & 0.1482 & 0.1883 & 0.1455 & 0.1957 & 0.1526 & 0.1900 & 0.1471 &8.6$\pm$4.0 & 13.8$\pm$6.2 & 5.6$\pm$5.4 &&0.1953 & 0.1913 & 0.1974 & 0.1926 &12.2$\pm$4.4 & 16.5$\pm$6.4 & 8.6$\pm$8.7\\
$V_{\rm disp,80\%}$ && 0.1463 & 0.1122 & 0.1444 & 0.1106 & 0.1499 & 0.1153 & 0.1457 & 0.1117 &6.2$\pm$4.1 & 11.4$\pm$6.3 & 4.3$\pm$5.0 &&0.1551 & 0.1533 & 0.1523 & 0.1531 &11.2$\pm$4.1 & 14.6$\pm$6.2 & 10.8$\pm$8.7\\
\cline{3-13}\cline{15-21}
$\phi$ && 0.1667 & 0.1266 & 0.1448 & 0.1096 & 0.1791 & 0.1393 & 0.1533 & 0.1164 &21.9$\pm$4.3 & 18.6$\pm$5.6 & 20.1$\pm$5.3 &&0.1748 & 0.1520 & 0.1916 & 0.1609 &21.1$\pm$4.2 & 20.2$\pm$6.9 & 22.0$\pm$8.7\\
$\psi_1$ && 0.1458 & 0.1112 & 0.1402 & 0.1064 & 0.1499 & 0.1148 & 0.1425 & 0.1084 &10.5$\pm$4.0 & 13.0$\pm$5.7 & 9.4$\pm$5.6 &&0.1506 & 0.1460 & 0.1514 & 0.1471 &11.4$\pm$4.4 & 13.8$\pm$6.8 & 12.6$\pm$9.2\\
$\psi_2$ && 0.1436 & 0.1099 & 0.1406 & 0.1073 & \cellcolor{blue!25}\textbf{0.1461} & \cellcolor{blue!25}\textbf{0.1121} & 0.1418 & 0.1084 &9.3$\pm$4.0 & 12.8$\pm$5.4 & 7.7$\pm$5.3 &&0.1548 & 0.1520 & 0.1517 & 0.1519 &12.8$\pm$4.4 & 15.8$\pm$6.4 & 12.6$\pm$8.8\\
$\psi_3$ && 0.1457 & 0.1109 & 0.1419 & 0.1077 & 0.1501 & 0.1146 & 0.1438 & 0.1093 &8.0$\pm$4.1 & 11.6$\pm$6.3 & 6.9$\pm$5.3 &&0.1515 & 0.1475 & 0.1523 & 0.1485 &10.6$\pm$4.4 & 12.6$\pm$6.6 & 11.8$\pm$8.9\\
$\psi_4$ && 0.1444 & 0.1104 & 0.1425 & 0.1088 & 0.1471 & 0.1128 & 0.1436 & 0.1097 &7.2$\pm$4.0 & 12.2$\pm$6.0 & 5.6$\pm$5.3 &&0.1568 & 0.1548 & 0.1532 & 0.1545 &11.7$\pm$4.3 & 14.8$\pm$6.7 & 12.1$\pm$9.0\\
$\psi_5$ && \cellcolor{blue!25}\textbf{0.1412} & \cellcolor{blue!25}\textbf{0.1080} & \cellcolor{blue!25}\textbf{0.1391} & \cellcolor{blue!25}\textbf{0.1063} & 0.1464 & 0.1128 & \cellcolor{blue!25}\textbf{0.1408} & \cellcolor{blue!25}\textbf{0.1077} &\cellcolor{blue!25}\textbf{3.7}$\pm$4.2 & 7.8$\pm$6.0 & 3.7$\pm$5.6 &&\cellcolor{blue!25}\textbf{0.1450} & \cellcolor{blue!25}\textbf{0.1436} & \cellcolor{blue!25}\textbf{0.1458} & \cellcolor{blue!25}\textbf{0.1440} &7.7$\pm$4.5 & 10.0$\pm$6.8 & 6.7$\pm$8.2\\
$\psi_6$ && 0.1423 & 0.1092 & 0.1407 & 0.1079 & 0.1463 & 0.1129 & 0.1420 & 0.1090 &4.3$\pm$4.1 & 8.8$\pm$5.9 & \cellcolor{blue!25}\textbf{3.3}$\pm$4.9 &&0.1510 & 0.1505 & 0.1495 & 0.1503 &8.6$\pm$4.4 & 11.6$\pm$5.9 & \cellcolor{blue!25}\textbf{6.5}$\pm$8.0\\
\hline
\hline
$M_{DM,peak}$&\multirow{17}{*}{\rotatebox[origin=c]{90}{redshift 0.5}}& 0.2096 & 0.1604 & 0.1945 & 0.1492 & 0.2144 & 0.1660 & 0.1981 & 0.1521 &20.0$\pm$5.1 & 20.0$\pm$6.9 & 18.5$\pm$11.1 &\multirow{17}{*}{\rotatebox[origin=c]{90}{redshift 2.0}}&0.2196 & 0.2093 & 0.2268 & 0.2116 &8.0$\pm$5.6 & 8.8$\pm$7.2 & 7.2$\pm$13.1 \\
$M_{\rm DM,90\%}$ && 0.2032 & 0.1527 & 0.1866 & 0.1396 & 0.2089 & 0.1608 & 0.1907 & 0.1433 &20.7$\pm$5.1 & 22.5$\pm$7.1 & 20.1$\pm$11.5 &&0.2066 & 0.1968 & 0.2171 & 0.1995 &8.5$\pm$5.8 & 9.0$\pm$6.4 & 7.4$\pm$12.8\\
$M_{\rm R_{max},peak}$ && 0.2607 & 0.2021 & 0.2528 & 0.1956 & 0.2566 & 0.1998 & 0.2535 & 0.1963 &20.9$\pm$5.3 & 22.6$\pm$7.3 & 13.2$\pm$11.9 &&0.2631 & 0.2570 & 0.2612 & 0.2575 &8.2$\pm$6.6 & 9.6$\pm$7.2 & 7.4$\pm$14.4\\
$M_{\rm R_{max},80\%}$ && 0.2428 & 0.1847 & 0.2356 & 0.1784 & 0.2436 & 0.1878 & 0.2370 & 0.1801 &17.3$\pm$5.1 & 19.5$\pm$6.9 & 13.7$\pm$10.7 &&0.2515 & 0.2485 & 0.2500 & 0.2487 &7.8$\pm$6.3 & 8.2$\pm$7.5 & 6.8$\pm$11.3\\
$|\dot{M_{\rm DM}}|_{\rm peak}$ && 0.4189 & 0.3362 & 0.4206 & 0.3369 & 0.4056 & 0.3297 & 0.4180 & 0.3356 &8.6$\pm$5.3 & 12.2$\pm$7.2 & 15.7$\pm$12.1 &&0.4002 & 0.4010 & 0.3903 & 0.3997 &4.2$\pm$5.8 & 6.1$\pm$7.0 & 22.6$\pm$17.7\\
$|\dot{M_{\rm DM}}|_{\rm 60\%}$ && 0.1988 & 0.1493 & 0.1975 & 0.1478 & 0.2016 & 0.1532 & 0.1982 & 0.1487 &\cellcolor{blue!25}\textbf{3.7}$\pm$5.1 & \cellcolor{blue!25}\textbf{4.3}$\pm$6.4 & 11.9$\pm$12.2 &&0.2164 & 0.2169 & 0.2130 & 0.2164 &\cellcolor{blue!25}\textbf{2.3}$\pm$5.4 & \cellcolor{blue!25}\textbf{2.8}$\pm$6.4 & \cellcolor{blue!25}\textbf{4.3}$\pm$13.8\\
$V_{\rm peak}$ && 0.1783 & 0.1371 & 0.1744 & 0.1341 & 0.1765 & 0.1363 & 0.1748 & 0.1345 &13.6$\pm$5.2 & 15.4$\pm$6.9 & 13.7$\pm$11.2 &&0.1948 & 0.1924 & 0.1874 & 0.1918 &6.9$\pm$6.1 & 8.0$\pm$7.4 & 5.0$\pm$12.5\\
$V_{\rm 90\%}$ && 0.1558 & 0.1153 & 0.1525 & 0.1121 & 0.1536 & 0.1159 & 0.1527 & 0.1127 &11.6$\pm$5.1 & 12.5$\pm$7.2 & 18.2$\pm$12.6 &&0.1757 & 0.1742 & 0.1657 & 0.1732 &6.7$\pm$6.1 & 7.6$\pm$7.2 & 6.6$\pm$14.3\\
$V_{\rm disp,peak}$ && 0.2007 & 0.1548 & 0.1972 & 0.1519 & 0.2001 & 0.1549 & 0.1977 & 0.1524 &13.8$\pm$5.0 & 16.2$\pm$7.1 & 11.2$\pm$11.5 &&0.2205 & 0.2184 & 0.2156 & 0.2180 &7.4$\pm$6.2 & 8.5$\pm$7.1 & 6.0$\pm$13.3\\
$V_{\rm disp,80\%}$ && 0.1663 & 0.1234 & 0.1648 & 0.1215 & 0.1586 & 0.1194 & 0.1638 & 0.1211 &12.0$\pm$5.3 & 14.3$\pm$7.5 & 16.8$\pm$11.1 &&0.1911 & 0.1911 & 0.1766 & 0.1893 &6.9$\pm$6.0 & 7.5$\pm$6.5 & 8.2$\pm$14.2\\
\cline{3-13}\cline{15-21}
$\phi$ && 0.1800 & 0.1333 & 0.1613 & 0.1212 & 0.1957 & 0.1493 & 0.1678 & 0.1261 &19.4$\pm$5.0 & 20.2$\pm$7.3 & 20.6$\pm$11.7 &&0.2013 & 0.1917 & 0.1955 & 0.1922 &10.2$\pm$5.9 & 10.9$\pm$6.9 & 8.5$\pm$12.8\\
$\psi_1$ && 0.1569 & 0.1162 & 0.1527 & 0.1122 & 0.1542 & 0.1165 & 0.1530 & 0.1130 &12.7$\pm$5.4 & 14.4$\pm$6.7 & 20.1$\pm$12.3 &&0.1762 & 0.1740 & 0.1672 & 0.1731 &6.7$\pm$6.2 & 7.2$\pm$6.4 & 7.4$\pm$13.4\\
$\psi_2$ && 0.1652 & 0.1230 & 0.1626 & 0.1202 & 0.1588 & 0.1197 & 0.1619 & 0.1201 &12.9$\pm$5.3 & 14.9$\pm$6.9 & 17.0$\pm$11.8 &&0.1816 & 0.1786 & 0.1757 & 0.1783 &7.6$\pm$6.4 & 8.5$\pm$7.4 & 8.1$\pm$13.3\\
$\psi_3$ && 0.1583 & 0.1175 & 0.1547 & 0.1140 & 0.1554 & 0.1174 & 0.1548 & 0.1146 &12.1$\pm$5.5 & 13.7$\pm$6.7 & 18.6$\pm$12.9 &&0.1780 & 0.1762 & 0.1691 & 0.1753 &6.5$\pm$5.8 & 7.1$\pm$6.8 & 6.9$\pm$13.9\\
$\psi_4$ && 0.1691 & 0.1256 & 0.1673 & 0.1234 & 0.1609 & 0.1213 & 0.1662 & 0.1230 &12.6$\pm$5.5 & 15.4$\pm$6.9 & 11.3$\pm$10.8 &&0.1940 & 0.1938 & 0.1800 & 0.1921 &6.8$\pm$6.3 & 7.4$\pm$7.1 & 7.0$\pm$13.3\\
$\psi_5$ && \cellcolor{blue!25}\textbf{0.1511} & \cellcolor{blue!25}\textbf{0.1121} & \cellcolor{blue!25}\textbf{0.1501} & \cellcolor{blue!25}\textbf{0.1107} & \cellcolor{blue!25}\textbf{0.1484} & \cellcolor{blue!25}\textbf{0.1123} & \cellcolor{blue!25}\textbf{0.1498} & \cellcolor{blue!25}\textbf{0.1110} &9.3$\pm$5.1 & 10.4$\pm$7.0 & \cellcolor{blue!25}\textbf{10.7}$\pm$12.1 &&\cellcolor{blue!25}\textbf{0.1697} & \cellcolor{blue!25}\textbf{0.1695} & \cellcolor{blue!25}\textbf{0.1610} & \cellcolor{blue!25}\textbf{0.1684} &4.9$\pm$5.9 & 5.5$\pm$6.9 & 4.9$\pm$13.4\\
$\psi_6$ && 0.1603 & 0.1190 & 0.1602 & 0.1184 & 0.1554 & 0.1175 & 0.1594 & 0.1183 &8.8$\pm$5.2 & 9.9$\pm$7.1 & 14.7$\pm$13.0 &&0.1800 & 0.1804 & 0.1713 & 0.1793 &4.8$\pm$6.1 & 5.1$\pm$6.6 & 5.0$\pm$12.1\
\enddata
\end{deluxetable*}
\end{rotatetable*}

\bibliography{main}
\appendix
\section{Scatter between Stellar Mass and Subhalo Properties}
\label{sec:appa}
In Section \ref{sec:rank}, we assigned model galaxies to subhalos strictly following the rank order (in $\psi$) to facilitate a direct comparison between our results and those of TO21. However, in the usual implementation of SHAM, a stellar mass function is used to assign $M_\star$ to subhalos of the same number density. In so doing, a scatter between dark matter subhalo properties ($X$) and $M_\star$ is found to be necessary. 
Effectively, our approach assumes zero scatter.  Here, we examine the effect of introducing a scatter on our results.
Specifically, we artificially introduce an uncertainty in stellar mass through the use of log-normal probability distributions with standard deviations of $\sigma_\star=0.1$ and $0.4$ dex for $M_\star$, respectively. Subsequently, we evaluate the loss metrics for all adopted AM schemes utilizing this uncertainty. To ensure the robustness of our results, this calculation was repeated 1000 times for all AM schemes and the averages of the loss metrics were subsequently reported.

\begin{figure*}
  \centering
  \includegraphics[width=1.0\linewidth]{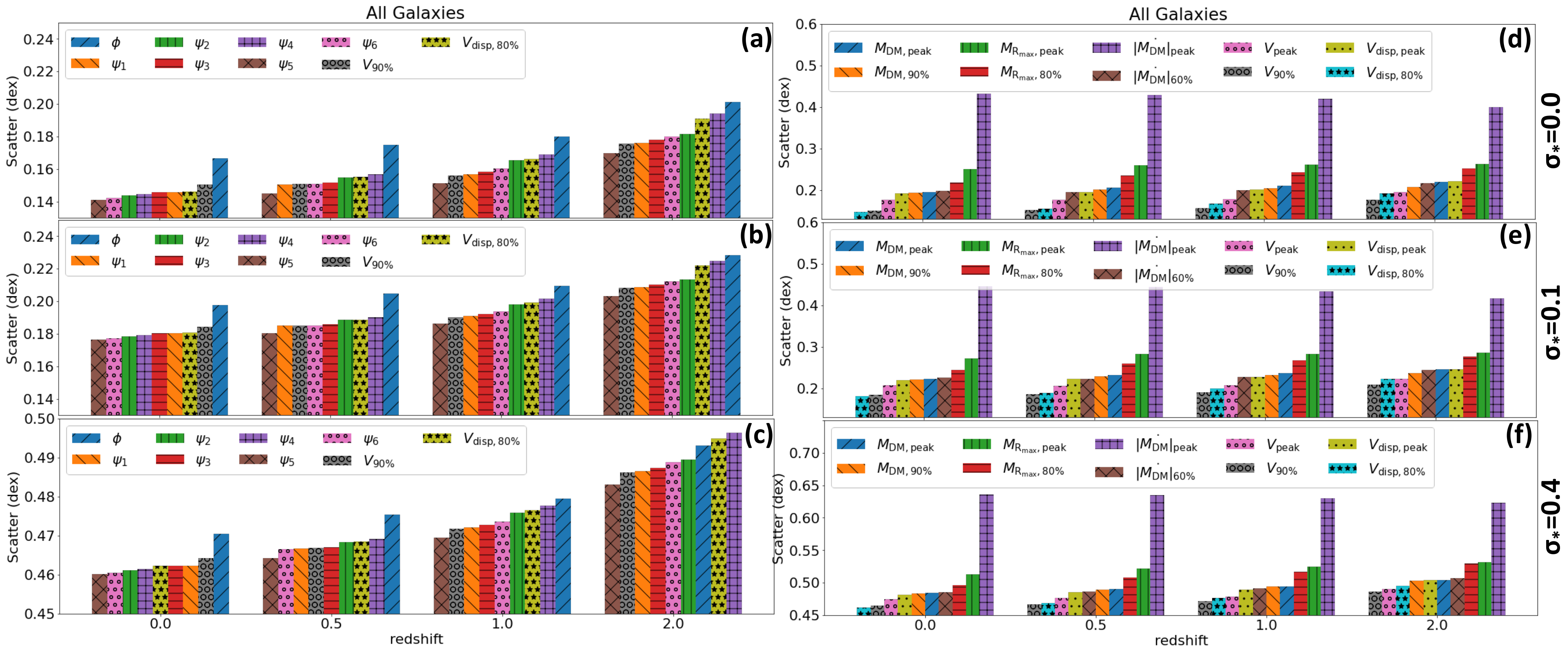}
  \caption{The panels located on the left and right sides of this figure depict the same data as that presented in Figure \ref{fig:scat}(h) and Figure \ref{fig:scat}(g), respectively. However, these representations differ in the log-normal scatter values that are indicated in the rightmost of each row of panels.}
  \label{fig:comp_scat}
\end{figure*}

\begin{figure*}
  \centering
  \includegraphics[width=1.0\linewidth]{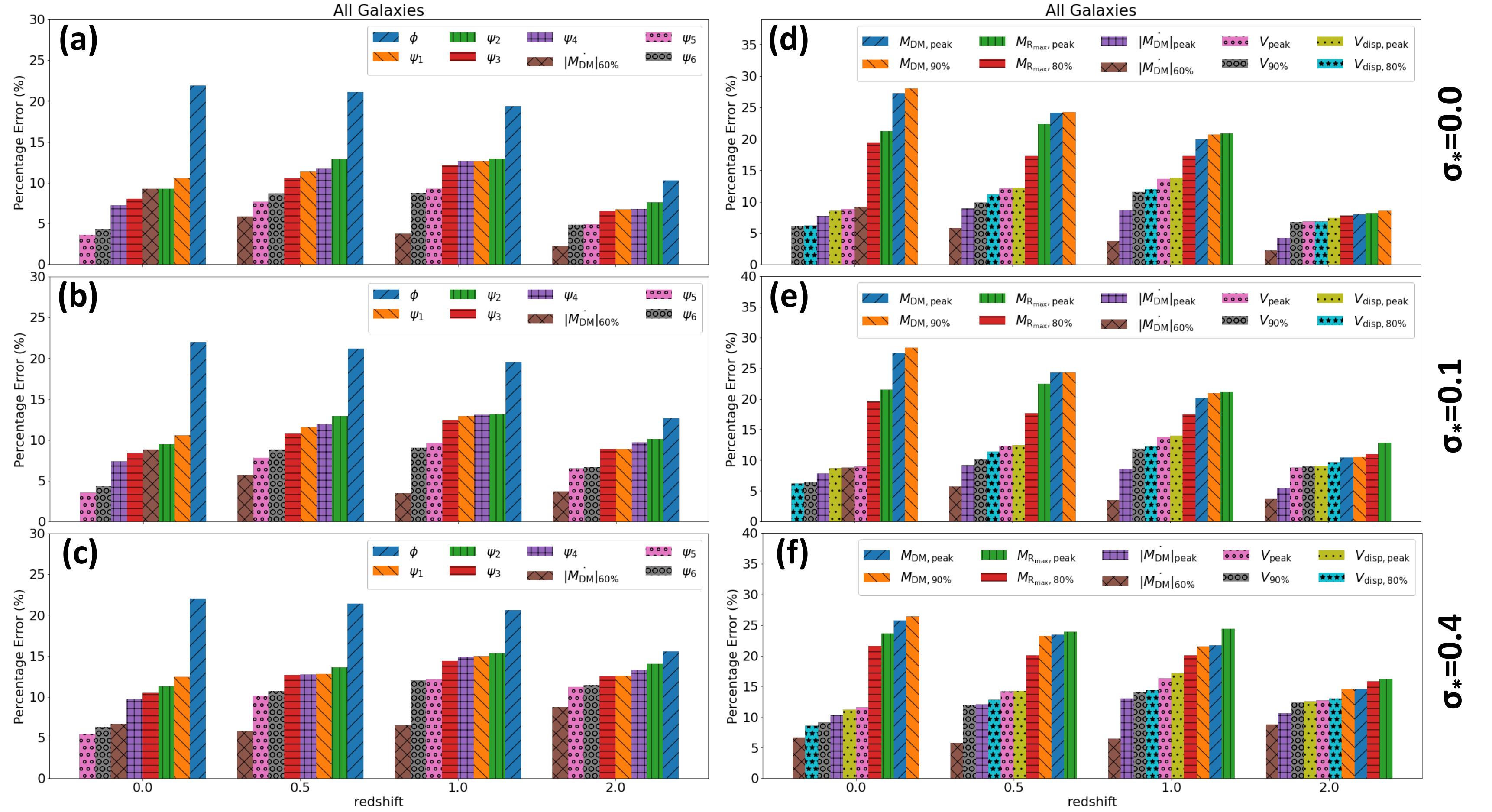}
  \caption{The panels located on the left and right sides of this figure depict the same data as that presented in Figure \ref{fig:clust_col}(h) and Figure \ref{fig:clust_col}(g), respectively. However, these representations differ in the log-normal scatter values that are indicated in the rightmost of each row of panels.}
  \label{fig:comp_clust}
\end{figure*}

\begin{figure*}
  \centering
  \includegraphics[width=1.0\linewidth]{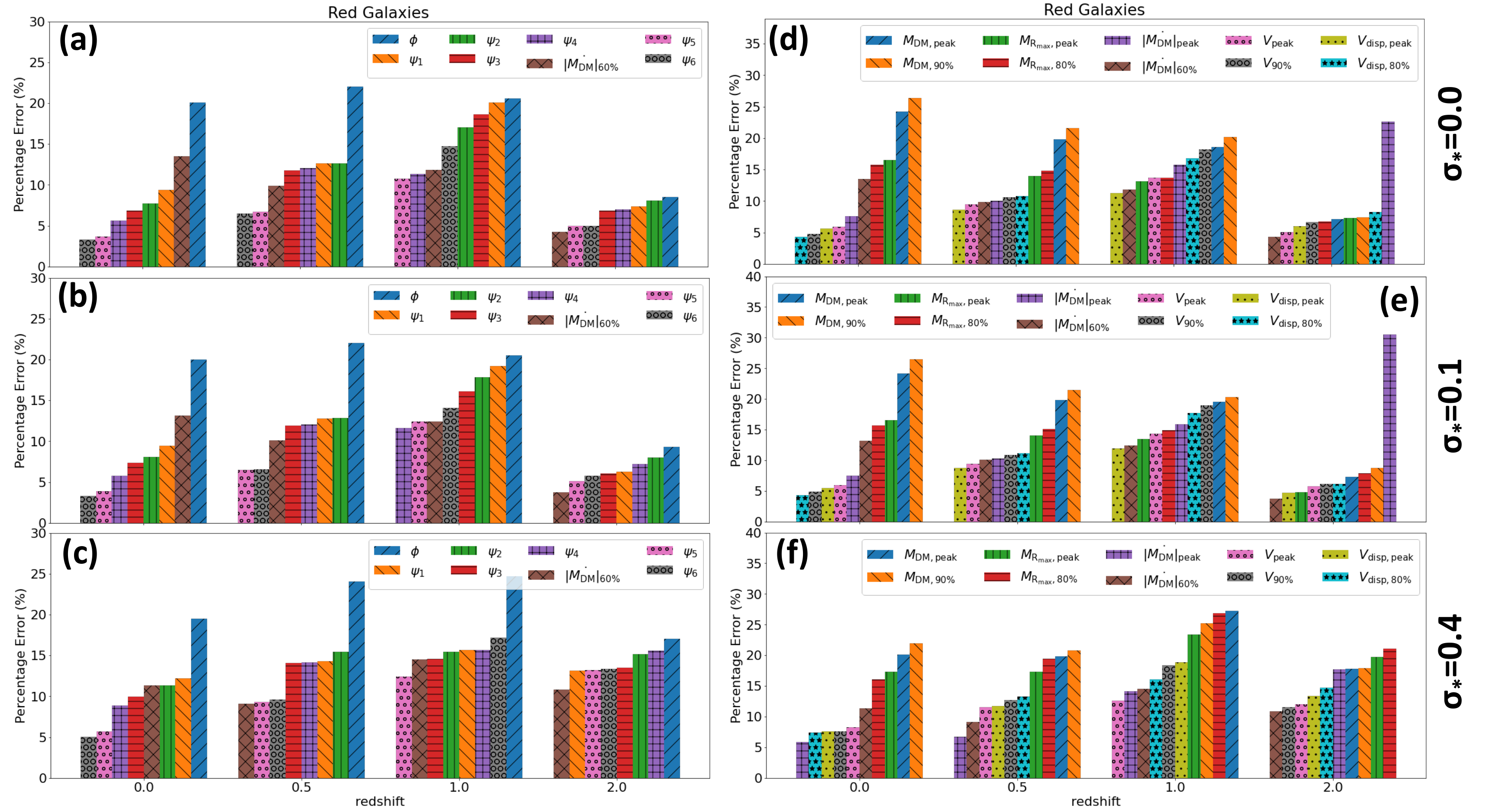}
  \caption{Same as Figure \ref{fig:comp_clust}, but with red galaxies.}
  \label{fig:comp_clust_red}
\end{figure*}

\begin{figure*}
  \centering
  \includegraphics[width=1.0\linewidth]{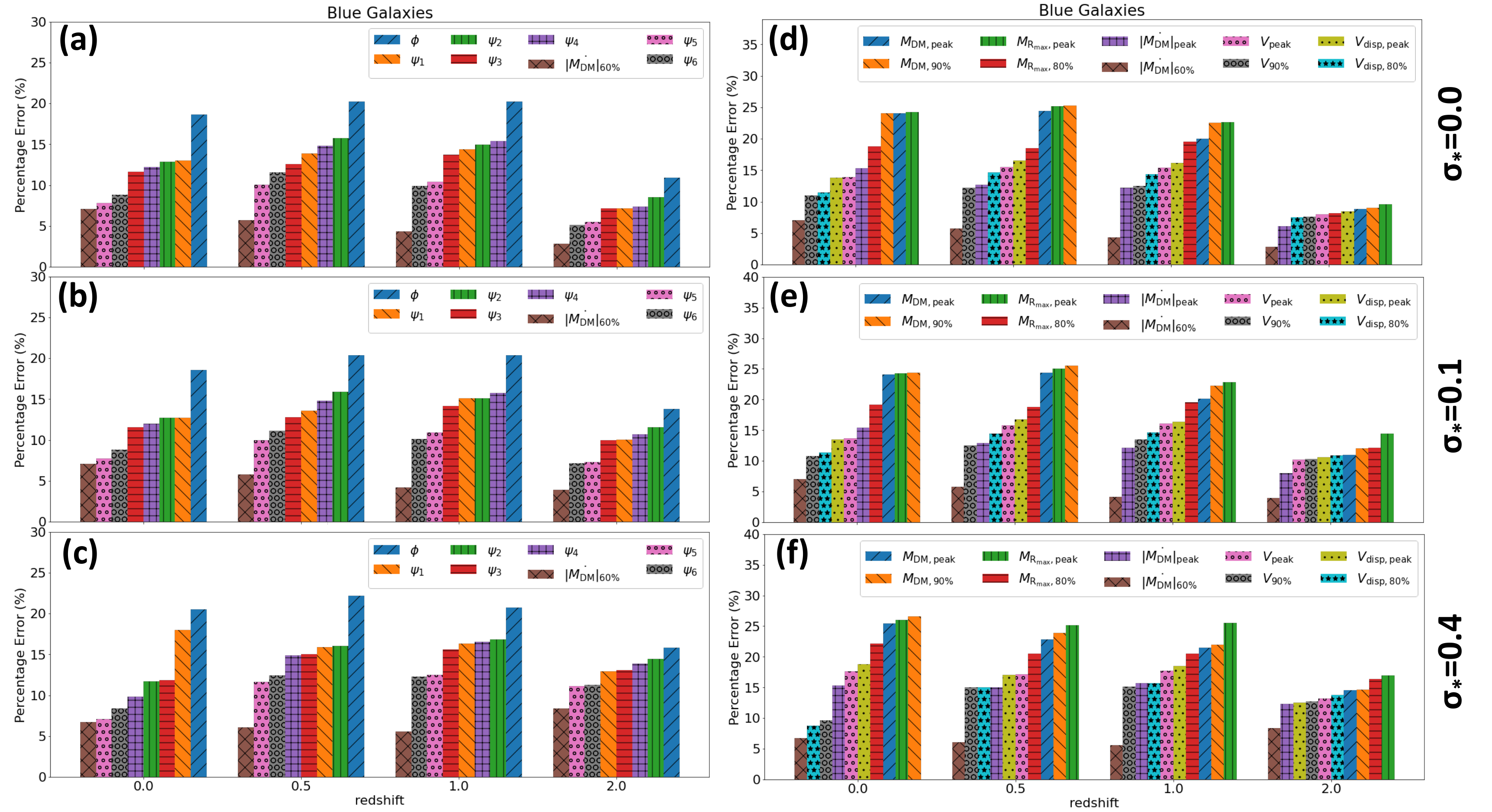}
  \caption{Same as Figure \ref{fig:comp_clust}, but with blue galaxies.}
  \label{fig:comp_clust_blue}
\end{figure*}

Figure \ref{fig:comp_scat}- \ref{fig:comp_clust_blue} present the same information as that of panels (g) and (h)  of Figure \ref{fig:scat}- \ref{fig:clust_red}, and panels (e) and (f) of Figure \ref{fig:clust_blue}, respectively, but with different $\sigma_\star$'s. Specifically, the figures compare the results obtained with $\sigma_\star=0.1$ and $0.4$ dex. It is noteworthy that the relative performances among the various AM schemes evaluated, as depicted in these figures, are overall consistent with the relative performances obtained in the absence of uncertainties ($\sigma_\star=0$).

\newpage

\movetabledown=7.5cm
\begin{rotatetable*}
\begin{deluxetable*}{||l||l|llllllll|lll||l|llll|lll}
\tabletypesize{\footnotesize}

\tablecaption{Same as Table \ref{tab:comp}, but with $\sigma_\star=0.1$ dex. \label{tab:comp1}}
\tablehead{
Loss Metrics& \multicolumn{1}{c}{}& \multicolumn{8}{c}{Stellar Mass Prediction} & \multicolumn{3}{c}{Prediction of $w_p$}& \multicolumn{1}{c}{}
& \multicolumn{4}{c}{Stellar Mass Prediction} & \multicolumn{3}{c}{Prediction of $w_p$}\\
\midrule
Galaxy Types&\multicolumn{1}{c}{}& \multicolumn{2}{c}{All}& \multicolumn{2}{c}{Central}& \multicolumn{2}{c}{Satellite}& \multicolumn{2}{c}{Mix}&All & Blue & \multicolumn{1}{c}{Red}&\multicolumn{1}{c}{}
& All& Central& Satellite& \multicolumn{1}{c}{Mix}& \multicolumn{1}{c}{All} & Blue & Red\\
\midrule
AM Scheme&\multicolumn{1}{c}{}& Scatter & Loss & Scatter & Loss & Scatter & Loss & Scatter & \multicolumn{1}{c}{Loss} & \multicolumn{3}{c}{Percentage Error}&\multicolumn{1}{c}{}
& \multicolumn{4}{c}{Scatter}& \multicolumn{3}{c}{Percentage Error}\\
}
\startdata
$M_{DM,peak}$& \multirow{17}{*}{\rotatebox[origin=c]{90}{redshift 0}}& 0.2225 & 0.1729 & 0.2055 & 0.1595 & 0.2211 & 0.1715 & 0.2092 & 0.1622 &27.4$\pm$4.1 & 24.1$\pm$6.2 & 24.2$\pm$5.6&\multirow{17}{*}{\rotatebox[origin=c]{90}{redshift 1.0}}&0.2316 & 0.2155 & 0.2338 & 0.2194 &24.3$\pm$4.3 & 24.3$\pm$6.8 & 19.9$\pm$8.9 \\
$M_{\rm DM,90\%}$ && 0.2205 & 0.1710 & 0.2004 & 0.1549 & 0.2190 & 0.1709 & 0.2048 & 0.1586 &28.3$\pm$4.0 & 24.4$\pm$5.5 & 26.5$\pm$5.7 &&0.2279 & 0.2100 & 0.2292 & 0.2141 &24.3$\pm$4.3 & 25.5$\pm$7.0 & 21.5$\pm$8.6\\
$M_{\rm R_{max},peak}$ && 0.2717 & 0.2116 & 0.2673 & 0.2084 & 0.2635 & 0.2045 & 0.2664 & 0.2075 &21.5$\pm$3.8 & 24.3$\pm$7.1 & 16.5$\pm$5.4 &&0.2819 & 0.2746 & 0.2764 & 0.2750 &22.5$\pm$4.3 & 25.0$\pm$6.6 & 14.0$\pm$8.5\\
$M_{\rm R_{max},80\%}$ && 0.2436 & 0.1890 & 0.2375 & 0.1842 & 0.2416 & 0.1875 & 0.2384 & 0.1850 &19.5$\pm$3.9 & 19.2$\pm$6.4 & 15.7$\pm$5.6 &&0.2588 & 0.2521 & 0.2595 & 0.2537 &17.6$\pm$4.3 & 18.8$\pm$6.8 & 15.1$\pm$8.1\\
$|\dot{M_{\rm DM}}|_{\rm peak}$ && 0.4458 & 0.3622 & 0.4522 & 0.3679 & 0.4213 & 0.3417 & 0.4453 & 0.3619 &7.8$\pm$3.8 & 15.4$\pm$7.6 & 7.5$\pm$5.8 &&0.4437 & 0.4467 & 0.4275 & 0.4429 &9.1$\pm$4.1 & 12.9$\pm$6.1 & 10.3$\pm$8.3\\
$|\dot{M_{\rm DM}}|_{\rm 60\%}$ && 0.2249 & 0.1754 & 0.2200 & 0.1712 & 0.2259 & 0.1761 & 0.2214 & 0.1723 &8.8$\pm$4.3 & \cellcolor{blue!25}\textbf{7.1}$\pm$5.9 & 13.1$\pm$5.4 &&0.2230 & 0.2212 & 0.2242 & 0.2218 &\cellcolor{blue!25}\textbf{5.7}$\pm$4.4 & \cellcolor{blue!25}\textbf{5.8}$\pm$5.9 & 10.1$\pm$8.4\\
$V_{\rm peak}$ && 0.2064 & 0.1618 & 0.2040 & 0.1599 & 0.2085 & 0.1637 & 0.2050 & 0.1608 &9.0$\pm$4.0 & 13.7$\pm$6.0 & 5.9$\pm$5.1 &&0.2061 & 0.2029 & 0.2059 & 0.2035 &12.3$\pm$4.1 & 15.8$\pm$6.3 & 9.4$\pm$8.7\\
$V_{\rm 90\%}$ && 0.1842 & 0.1437 & 0.1816 & 0.1416 & 0.1885 & 0.1473 & 0.1832 & 0.1429 &6.3$\pm$3.9 & 10.8$\pm$5.6 & 4.9$\pm$5.2 &&0.1852 & 0.1825 & 0.1855 & 0.1831 &10.1$\pm$4.5 & 12.5$\pm$6.8 & 10.8$\pm$8.5\\
$V_{\rm disp,peak}$ && 0.2190 & 0.1721 & 0.2165 & 0.1701 & 0.2224 & 0.1750 & 0.2179 & 0.1712 &8.7$\pm$4.0 & 13.4$\pm$6.2 & 5.4$\pm$5.4 &&0.2226 & 0.2192 & 0.2240 & 0.2202 &12.5$\pm$4.4 & 16.8$\pm$6.3 & 8.8$\pm$8.7\\
$V_{\rm disp,80\%}$ && 0.1807 & 0.1413 & 0.1794 & 0.1403 & 0.1829 & 0.1432 & 0.1802 & 0.1409 &6.2$\pm$4.1 & 11.4$\pm$6.2 & 4.3$\pm$4.9 &&0.1888 & 0.1876 & 0.1854 & 0.1872 &11.4$\pm$4.1 & 14.4$\pm$6.1 & 11.1$\pm$8.8\\
\cline{3-13}\cline{15-21}
$\phi$ && 0.1977 & 0.1538 & 0.1797 & 0.1399 & 0.2078 & 0.1633 & 0.1865 & 0.1452 &22.0$\pm$4.3 & 18.5$\pm$5.7 & 20.0$\pm$5.3 &&0.2047 & 0.1856 & 0.2191 & 0.1929 &21.2$\pm$4.2 & 20.4$\pm$7.1 & 22.0$\pm$8.8\\
$\psi_1$ && 0.1804 & 0.1407 & 0.1762 & 0.1371 & 0.1829 & 0.1429 & 0.1777 & 0.1385 &10.6$\pm$4.0 & 12.7$\pm$5.7 & 9.5$\pm$5.6 &&0.1850 & 0.1815 & 0.1847 & 0.1822 &11.6$\pm$4.4 & 13.6$\pm$6.7 & 12.8$\pm$9.2\\
$\psi_2$ && 0.1785 & 0.1395 & 0.1763 & 0.1377 & \cellcolor{blue!25}\textbf{0.1797} & \cellcolor{blue!25}\textbf{0.1406} & 0.1771 & 0.1383 &9.5$\pm$3.9 & 12.7$\pm$5.5 & 8.1$\pm$5.3 &&0.1884 & 0.1865 & 0.1849 & 0.1862 &13.0$\pm$4.4 & 15.9$\pm$6.3 & 12.8$\pm$8.7\\
$\psi_3$ && 0.1803 & 0.1405 & 0.1775 & 0.1382 & 0.1831 & 0.1428 & 0.1788 & 0.1392 &8.4$\pm$4.1 & 11.6$\pm$6.3 & 7.3$\pm$5.3 &&0.1857 & 0.1828 & 0.1855 & 0.1834 &10.8$\pm$4.3 & 12.8$\pm$6.5 & 11.9$\pm$8.9\\
$\psi_4$ && 0.1792 & 0.1399 & 0.1779 & 0.1389 & 0.1806 & 0.1411 & 0.1785 & 0.1394 &7.4$\pm$4.0 & 12.0$\pm$6.0 & 5.8$\pm$5.3 &&0.1902 & 0.1888 & 0.1862 & 0.1883 &11.9$\pm$4.3 & 14.8$\pm$6.7 & 12.0$\pm$9.0\\
$\psi_5$ && \cellcolor{blue!25}\textbf{0.1765} & \cellcolor{blue!25}\textbf{0.1380} & \cellcolor{blue!25}\textbf{0.1751} & \cellcolor{blue!25}\textbf{0.1368} & 0.1799 & 0.1410 & \cellcolor{blue!25}\textbf{0.1762} & \cellcolor{blue!25}\textbf{0.1378} &\cellcolor{blue!25}\textbf{3.5}$\pm$4.2 & 7.7$\pm$6.0 & 3.9$\pm$5.6 &&\cellcolor{blue!25}\textbf{0.1804} & \cellcolor{blue!25}\textbf{0.1795} & \cellcolor{blue!25}\textbf{0.1800} & \cellcolor{blue!25}\textbf{0.1796} &7.8$\pm$4.5 & 10.0$\pm$6.8 & \cellcolor{blue!25}\textbf{6.5}$\pm$8.2\\
$\psi_6$ && 0.1774 & 0.1388 & 0.1764 & 0.1380 & 0.1798 & 0.1410 & 0.1772 & 0.1387 &4.3$\pm$4.1 & 8.8$\pm$6.0 & \cellcolor{blue!25}\textbf{3.3}$\pm$5.1 &&0.1852 & 0.1852 & 0.1830 & 0.1848 &8.8$\pm$4.4 & 11.1$\pm$6.9 & 6.6$\pm$8.1\\
\hline
\hline
$M_{DM,peak}$&\multirow{17}{*}{\rotatebox[origin=c]{90}{redshift 0.5}}& 0.2356 & 0.1839 & 0.2222 & 0.1738 & 0.2396 & 0.1876 & 0.2253 & 0.1762 &20.1$\pm$5.1 & 20.2$\pm$6.9 & 19.5$\pm$11.1 &\multirow{17}{*}{\rotatebox[origin=c]{90}{redshift 2.0}}&0.2450 & 0.2357 & 0.2515 & 0.2377 &10.4$\pm$5.5 & 11.0$\pm$6.1 & 7.3$\pm$13.4 \\
$M_{\rm DM,90\%}$ && 0.2312 & 0.1773 & 0.2169 & 0.1657 & 0.2353 & 0.1831 & 0.2202 & 0.1687 &21.0$\pm$5.1 & 22.2$\pm$7.3 & 20.3$\pm$11.2 &&0.2356 & 0.2273 & 0.2438 & 0.2294 &10.5$\pm$5.8 & 12.0$\pm$6.4 & 8.7$\pm$13.5\\
$M_{\rm R_{max},peak}$ && 0.2828 & 0.2220 & 0.2755 & 0.2162 & 0.2789 & 0.2189 & 0.2761 & 0.2167 &21.2$\pm$5.3 & 22.9$\pm$7.2 & 13.5$\pm$11.6 &&0.2856 & 0.2799 & 0.2838 & 0.2804 &12.8$\pm$6.1 & 14.4$\pm$6.5 & 4.8$\pm$13.6\\
$M_{\rm R_{max},80\%}$ && 0.2672 & 0.2062 & 0.2609 & 0.2008 & 0.2673 & 0.2080 & 0.2620 & 0.2020 &17.5$\pm$5.1 & 19.6$\pm$6.9 & 14.9$\pm$10.7 &&0.2767 & 0.2741 & 0.2742 & 0.2741 &11.0$\pm$6.1 & 12.1$\pm$6.5 & 7.9$\pm$12.9\\
$|\dot{M_{\rm DM}}|_{\rm peak}$ && 0.4336 & 0.3495 & 0.4352 & 0.3504 & 0.4212 & 0.3425 & 0.4328 & 0.3490 &8.6$\pm$5.2 & 12.1$\pm$7.1 & 15.9$\pm$12.2 &&0.4162 & 0.4171 & 0.4070 & 0.4158 &5.4$\pm$5.5 & 8.0$\pm$5.9 & 30.5$\pm$14.4\\
$|\dot{M_{\rm DM}}|_{\rm 60\%}$ && 0.2269 & 0.1739 & 0.2260 & 0.1728 & 0.2288 & 0.1766 & 0.2265 & 0.1735 &\cellcolor{blue!25}\textbf{3.5}$\pm$5.0 & \cellcolor{blue!25}\textbf{4.2}$\pm$6.7 & 12.4$\pm$12.1 &&0.2443 & 0.2449 & 0.2400 & 0.2443 &\cellcolor{blue!25}\textbf{3.7}$\pm$5.8 & \cellcolor{blue!25}\textbf{3.9}$\pm$6.1 & \cellcolor{blue!25}\textbf{3.7}$\pm$12.4\\
$V_{\rm peak}$ && 0.2078 & 0.1629 & 0.2046 & 0.1605 & 0.2058 & 0.1613 & 0.2048 & 0.1606 &13.8$\pm$5.2 & 16.1$\pm$7.1 & 14.4$\pm$11.3 &&0.2224 & 0.2204 & 0.2155 & 0.2198 &8.8$\pm$5.9 & 10.2$\pm$6.3 & 5.8$\pm$13.5\\
$V_{\rm 90\%}$ && 0.1901 & 0.1448 & 0.1877 & 0.1424 & 0.1870 & 0.1441 & 0.1876 & 0.1427 &11.8$\pm$5.0 & 13.5$\pm$7.3 & 18.9$\pm$12.3 &&0.2083 & 0.2072 & 0.1982 & 0.2061 &9.0$\pm$6.3 & 10.3$\pm$6.8 & 6.1$\pm$14.0\\
$V_{\rm disp,peak}$ && 0.2275 & 0.1785 & 0.2245 & 0.1761 & 0.2265 & 0.1776 & 0.2248 & 0.1764 &14.0$\pm$5.0 & 16.3$\pm$7.0 & 11.9$\pm$11.5 &&0.2455 & 0.2437 & 0.2408 & 0.2433 &9.1$\pm$6.0 & 10.6$\pm$6.9 & 4.7$\pm$13.6\\
$V_{\rm disp,80\%}$ && 0.1991 & 0.1515 & 0.1981 & 0.1502 & 0.1914 & 0.1471 & 0.1969 & 0.1497 &12.2$\pm$5.2 & 14.6$\pm$7.4 & 17.7$\pm$11.1 &&0.2221 & 0.2224 & 0.2080 & 0.2206 &9.6$\pm$6.1 & 10.9$\pm$6.7 & 6.2$\pm$14.3\\
\cline{3-13}\cline{15-21}
$\phi$ && 0.2093 & 0.1606 & 0.1934 & 0.1498 & 0.2228 & 0.1731 & 0.1988 & 0.1539 &19.5$\pm$5.0 & 20.4$\pm$7.1 & 20.5$\pm$11.2 &&0.2282 & 0.2197 & 0.2229 & 0.2201 &12.6$\pm$6.5 & 13.8$\pm$6.9 & 9.3$\pm$13.7\\
$\psi_1$ && 0.1910 & 0.1455 & 0.1879 & 0.1425 & 0.1875 & 0.1446 & 0.1878 & 0.1428 &12.9$\pm$5.4 & 15.1$\pm$7.1 & 19.2$\pm$12.5 &&0.2087 & 0.2070 & 0.1995 & 0.2061 &8.9$\pm$6.1 & 10.0$\pm$7.1 & 6.3$\pm$13.8\\
$\psi_2$ && 0.1981 & 0.1511 & 0.1961 & 0.1490 & 0.1915 & 0.1473 & 0.1953 & 0.1487 &13.2$\pm$5.3 & 15.1$\pm$6.9 & 17.8$\pm$11.4 &&0.2135 & 0.2112 & 0.2070 & 0.2107 &10.1$\pm$6.3 & 11.6$\pm$6.8 & 8.0$\pm$12.2\\
$\psi_3$ && 0.1922 & 0.1465 & 0.1895 & 0.1439 & 0.1885 & 0.1454 & 0.1893 & 0.1441 &12.4$\pm$5.5 & 14.1$\pm$6.6 & 16.1$\pm$12.2 &&0.2103 & 0.2090 & 0.2011 & 0.2080 &8.9$\pm$6.4 & 10.0$\pm$6.6 & 6.1$\pm$13.4\\
$\psi_4$ && 0.2014 & 0.1534 & 0.2001 & 0.1518 & 0.1934 & 0.1487 & 0.1990 & 0.1513 &13.1$\pm$5.5 & 15.7$\pm$6.9 & \cellcolor{blue!25}\textbf{11.6}$\pm$10.6 &&0.2247 & 0.2248 & 0.2109 & 0.2231 &9.7$\pm$6.1 & 10.7$\pm$6.6 & 7.2$\pm$12.9\\
$\psi_5$ && \cellcolor{blue!25}\textbf{0.1861} & \cellcolor{blue!25}\textbf{0.1419} & \cellcolor{blue!25}\textbf{0.1855} & \cellcolor{blue!25}\textbf{0.1411} & \cellcolor{blue!25}\textbf{0.1826} & \cellcolor{blue!25}\textbf{0.1410} & \cellcolor{blue!25}\textbf{0.1850} & \cellcolor{blue!25}\textbf{0.1411} &9.6$\pm$5.1 & 10.9$\pm$7.0 & 12.4$\pm$12.2 &&\cellcolor{blue!25}\textbf{0.2030} & \cellcolor{blue!25}\textbf{0.2030} & \cellcolor{blue!25}\textbf{0.1941} & \cellcolor{blue!25}\textbf{0.2020} &6.5$\pm$6.7 & 7.3$\pm$7.1 & 5.1$\pm$13.2\\
$\psi_6$ && 0.1938 & 0.1477 & 0.1940 & 0.1475 & 0.1887 & 0.1454 & 0.1931 & 0.1471 &9.0$\pm$5.2 & 10.1$\pm$7.2 & 14.0$\pm$12.5 &&0.2123 & 0.2129 & 0.2032 & 0.2117 &6.6$\pm$6.1 & 7.2$\pm$6.3 & 5.8$\pm$12.7\\
\enddata
\end{deluxetable*}
\end{rotatetable*}

\movetabledown=7.5cm
\begin{rotatetable*}
\begin{deluxetable*}{||l||l|llllllll|lll||l|llll|lll}
\tabletypesize{\footnotesize}

\tablecaption{Same as Table \ref{tab:comp}, but with $\sigma_\star=0.4$ dex. \label{tab:comp2}}
\tablehead{
Loss Metrics& \multicolumn{1}{c}{}& \multicolumn{8}{c}{Stellar Mass Prediction} & \multicolumn{3}{c}{Prediction of $w_p$}& \multicolumn{1}{c}{}
& \multicolumn{4}{c}{Stellar Mass Prediction} & \multicolumn{3}{c}{Prediction of $w_p$}\\
\midrule
Galaxy Types&\multicolumn{1}{c}{}& \multicolumn{2}{c}{All}& \multicolumn{2}{c}{Central}& \multicolumn{2}{c}{Satellite}& \multicolumn{2}{c}{Mix}&All & Blue & \multicolumn{1}{c}{Red}&\multicolumn{1}{c}{}
& All& Central& Satellite& \multicolumn{1}{c}{Mix}& \multicolumn{1}{c}{All} & Blue & Red\\
\midrule
AM Scheme&\multicolumn{1}{c}{}& Scatter & Loss & Scatter & Loss & Scatter & Loss & Scatter & \multicolumn{1}{c}{Loss} & \multicolumn{3}{c}{Percentage Error}&\multicolumn{1}{c}{}
& \multicolumn{4}{c}{Scatter}& \multicolumn{3}{c}{Percentage Error}\\
}
\startdata
$M_{DM,peak}$& \multirow{17}{*}{\rotatebox[origin=c]{90}{redshift 0}}& 0.4840 & 0.3841 & 0.4759 & 0.3773 & 0.4818 & 0.3825 & 0.4772 & 0.3785 &25.7$\pm$4.0 & 25.5$\pm$7.8 & 20.1$\pm$5.4&\multirow{17}{*}{\rotatebox[origin=c]{90}{redshift 1.0}}&0.4904 & 0.4826 & 0.4902 & 0.4842 &23.4$\pm$4.2 & 22.9$\pm$6.0 & 19.8$\pm$9.4 \\
$M_{\rm DM,90\%}$ && 0.4835 & 0.3832 & 0.4738 & 0.3751 & 0.4811 & 0.3819 & 0.4755 & 0.3766 &26.4$\pm$3.8 & 26.6$\pm$7.4 & 22.0$\pm$5.3 &&0.4896 & 0.4811 & 0.4881 & 0.4825 &23.2$\pm$4.1 & 23.9$\pm$6.6 & 20.8$\pm$8.9\\
$M_{\rm R_{max},peak}$ && 0.5130 & 0.4080 & 0.5112 & 0.4065 & 0.5076 & 0.4033 & 0.5103 & 0.4058 &23.6$\pm$3.8 & 26.0$\pm$7.6 & 17.4$\pm$5.4 &&0.5216 & 0.5181 & 0.5173 & 0.5180 &23.9$\pm$4.2 & 25.2$\pm$7.0 & 17.4$\pm$8.9\\
$M_{\rm R_{max},80\%}$ && 0.4963 & 0.3940 & 0.4938 & 0.3918 & 0.4943 & 0.3925 & 0.4939 & 0.3920 &21.6$\pm$3.8 & 22.1$\pm$7.3 & 16.1$\pm$5.2 &&0.5083 & 0.5055 & 0.5070 & 0.5058 &20.0$\pm$4.2 & 20.5$\pm$6.6 & 19.5$\pm$8.6\\
$|\dot{M_{\rm DM}}|_{\rm peak}$ && 0.6355 & 0.5133 & 0.6398 & 0.5171 & 0.6213 & 0.5011 & 0.6356 & 0.5134 &10.3$\pm$3.8 & 15.3$\pm$7.9 & 5.8$\pm$5.7 &&0.6350 & 0.6368 & 0.6270 & 0.6348 &12.1$\pm$4.0 & 15.0$\pm$6.4 & \cellcolor{blue!25}\textbf{6.7}$\pm$7.8\\
$|\dot{M_{\rm DM}}|_{\rm 60\%}$ && 0.4853 & 0.3850 & 0.4840 & 0.3837 & 0.4850 & 0.3850 & 0.4842 & 0.3840 &6.6$\pm$4.1 & \cellcolor{blue!25}\textbf{6.7}$\pm$6.8 & 11.3$\pm$5.6 &&0.4866 & 0.4868 & 0.4856 & 0.4866 &\cellcolor{blue!25}\textbf{5.8}$\pm$4.0 & \cellcolor{blue!25}\textbf{6.1}$\pm$6.1 & 9.1$\pm$8.5\\
$V_{\rm peak}$ && 0.4750 & 0.3772 & 0.4749 & 0.3770 & 0.4749 & 0.3775 & 0.4749 & 0.3771 &11.5$\pm$3.9 & 17.7$\pm$7.5 & 8.2$\pm$5.3 &&0.4763 & 0.4757 & 0.4747 & 0.4755 &14.1$\pm$4.0 & 17.2$\pm$6.2 & 11.6$\pm$8.4\\
$V_{\rm 90\%}$ && 0.4643 & 0.3679 & 0.4642 & 0.3676 & 0.4648 & 0.3688 & 0.4644 & 0.3679 &9.2$\pm$3.8 & 9.7$\pm$5.8 & 7.6$\pm$5.3 &&0.4668 & 0.4667 & 0.4648 & 0.4663 &12.0$\pm$4.3 & 15.0$\pm$7.1 & 12.7$\pm$9.1\\
$V_{\rm disp,peak}$ && 0.4817 & 0.3828 & 0.4814 & 0.3825 & 0.4825 & 0.3837 & 0.4817 & 0.3828 &11.2$\pm$4.0 & 18.8$\pm$7.2 & 7.6$\pm$5.7 &&0.4852 & 0.4844 & 0.4846 & 0.4844 &14.2$\pm$4.2 & 17.0$\pm$6.2 & 11.7$\pm$9.2\\
$V_{\rm disp,80\%}$ && 0.4623 & 0.3663 & 0.4628 & 0.3666 & 0.4618 & 0.3665 & 0.4626 & 0.3666 &8.6$\pm$4.0 & 8.7$\pm$5.6 & 7.4$\pm$5.2 &&0.4685 & 0.4691 & 0.4647 & 0.4682 &12.9$\pm$3.9 & 15.0$\pm$6.3 & 13.3$\pm$8.8\\
\cline{3-13}\cline{15-21}
$\phi$ && 0.4705 & 0.3732 & 0.4627 & 0.3667 & 0.4746 & 0.3769 & 0.4655 & 0.3690 &21.9$\pm$4.1 & 20.5$\pm$7.8 & 19.5$\pm$5.4 &&0.4754 & 0.4669 & 0.4818 & 0.4700 &21.4$\pm$4.2 & 22.2$\pm$7.4 & 24.0$\pm$8.9\\
$\psi_1$ && 0.4623 & 0.3663 & 0.4615 & 0.3654 & 0.4620 & 0.3665 & 0.4616 & 0.3656 &12.4$\pm$3.8 & 18.0$\pm$7.5 & 12.2$\pm$5.8 &&0.4667 & 0.4662 & 0.4643 & 0.4658 &12.8$\pm$4.3 & 15.9$\pm$7.1 & 14.3$\pm$9.4\\
$\psi_2$ && 0.4612 & 0.3655 & 0.4612 & 0.3654 & 0.4602 & \cellcolor{blue!25}\textbf{0.3652} & 0.4610 & 0.3653 &11.3$\pm$3.9 & 11.7$\pm$5.9 & 11.3$\pm$5.5 &&0.4683 & 0.4685 & 0.4644 & 0.4677 &13.6$\pm$4.2 & 16.0$\pm$6.3 & 15.4$\pm$9.0\\
$\psi_3$ && 0.4623 & 0.3662 & 0.4621 & 0.3659 & 0.4621 & 0.3666 & 0.4621 & 0.3661 &10.5$\pm$4.0 & 11.8$\pm$6.9 & 9.9$\pm$5.5 &&0.4670 & 0.4668 & 0.4647 & 0.4664 &12.6$\pm$4.3 & 15.0$\pm$6.7 & 14.0$\pm$9.5\\
$\psi_4$ && 0.4615 & 0.3657 & 0.4620 & 0.3660 & 0.4607 & 0.3656 & 0.4617 & 0.3659 &9.7$\pm$4.0 & 9.8$\pm$6.1 & 8.9$\pm$5.6 &&0.4691 & 0.4696 & 0.4650 & 0.4687 &12.8$\pm$4.1 & 14.8$\pm$6.8 & 14.2$\pm$9.5\\
$\psi_5$ && \cellcolor{blue!25}\textbf{0.4602} & \cellcolor{blue!25}\textbf{0.3646} & \cellcolor{blue!25}\textbf{0.4607} & \cellcolor{blue!25}\textbf{0.3649} & 0.4603 & 0.3653 & \cellcolor{blue!25}\textbf{0.4606} & \cellcolor{blue!25}\textbf{0.3650} &\cellcolor{blue!25}\textbf{5.4}$\pm$4.0 & 7.1$\pm$6.3 & 5.7$\pm$5.4 &&\cellcolor{blue!25}\textbf{0.4643} & \cellcolor{blue!25}\textbf{0.4650} & \cellcolor{blue!25}\textbf{0.4619} & \cellcolor{blue!25}\textbf{0.4644} &10.1$\pm$4.4 & 11.6$\pm$6.6 & 9.3$\pm$8.5\\
$\psi_6$ && 0.4605 & 0.3649 & 0.4611 & 0.3653 & \cellcolor{blue!25}\textbf{0.4602} & 0.3652 & 0.4609 & 0.3653 &6.3$\pm$4.0 & 8.4$\pm$6.5 & \cellcolor{blue!25}\textbf{5.1}$\pm$5.0 &&0.4666 & 0.4678 & 0.4633 & 0.4668 &10.7$\pm$4.1 & 12.5$\pm$6.3 & 9.6$\pm$8.1\\
\hline
\hline
$M_{DM,peak}$&\multirow{17}{*}{\rotatebox[origin=c]{90}{redshift 0.5}}& 0.4946 & 0.3931 & 0.4878 & 0.3875 & 0.4954 & 0.3935 & 0.4892 & 0.3886 &21.6$\pm$4.9 & 21.5$\pm$6.1 & 27.3$\pm$10.3 &\multirow{17}{*}{\rotatebox[origin=c]{90}{redshift 2.0}}&0.5044 & 0.4992 & 0.5070 & 0.5002 &14.6$\pm$5.3 & 14.6$\pm$5.7 & 17.8$\pm$13.3 \\
$M_{\rm DM,90\%}$ && 0.4944 & 0.3901 & 0.4874 & 0.3841 & 0.4941 & 0.3912 & 0.4886 & 0.3853 &21.5$\pm$4.7 & 21.9$\pm$6.3 & 25.3$\pm$8.7 &&0.5028 & 0.4987 & 0.5043 & 0.4994 &14.5$\pm$5.1 & 14.6$\pm$5.8 & 17.9$\pm$13.2\\
$M_{\rm R_{max},peak}$ && 0.5248 & 0.4177 & 0.5210 & 0.4146 & 0.5214 & 0.4143 & 0.5211 & 0.4145 &24.4$\pm$5.1 & 25.5$\pm$6.7 & 23.4$\pm$11.5 &&0.5316 & 0.5284 & 0.5293 & 0.5285 &16.2$\pm$6.0 & 16.9$\pm$6.1 & 19.7$\pm$14.0\\
$M_{\rm R_{max},80\%}$ && 0.5166 & 0.4083 & 0.5138 & 0.4057 & 0.5146 & 0.4076 & 0.5139 & 0.4060 &20.0$\pm$4.9 & 20.5$\pm$6.5 & 26.9$\pm$11.1 &&0.5292 & 0.5284 & 0.5242 & 0.5279 &15.8$\pm$5.8 & 16.3$\pm$6.5 & 21.0$\pm$13.3\\
$|\dot{M_{\rm DM}}|_{\rm peak}$ && 0.6300 & 0.5068 & 0.6309 & 0.5075 & 0.6241 & 0.5031 & 0.6297 & 0.5067 &13.0$\pm$5.0 & 15.7$\pm$6.8 & 14.1$\pm$12.2 &&0.6232 & 0.6237 & 0.6183 & 0.6230 &10.6$\pm$5.8 & 12.3$\pm$5.9 & 17.7$\pm$16.5\\
$|\dot{M_{\rm DM}}|_{\rm 60\%}$ && 0.4910 & 0.3876 & 0.4914 & 0.3878 & 0.4902 & 0.3877 & 0.4912 & 0.3878 &\cellcolor{blue!25}\textbf{6.5}$\pm$5.0 & \cellcolor{blue!25}\textbf{5.6}$\pm$6.5 & 14.5$\pm$11.4 &&0.5069 & 0.5080 & 0.5012 & 0.5072 &\cellcolor{blue!25}\textbf{8.7}$\pm$5.8 & \cellcolor{blue!25}\textbf{8.3}$\pm$6.4 & \cellcolor{blue!25}\textbf{10.8}$\pm$13.3\\
$V_{\rm peak}$ && 0.4790 & 0.3805 & 0.4782 & 0.3798 & 0.4760 & 0.3781 & 0.4778 & 0.3795 &16.3$\pm$4.9 & 17.7$\pm$6.6 & 12.6$\pm$10.1 &&0.4903 & 0.4897 & 0.4846 & 0.4891 &12.7$\pm$5.5 & 13.2$\pm$6.0 & 12.0$\pm$13.6\\
$V_{\rm 90\%}$ && 0.4718 & 0.3717 & 0.4717 & 0.3713 & 0.4673 & 0.3694 & 0.4709 & 0.3710 &14.1$\pm$5.0 & 15.1$\pm$7.1 & 18.4$\pm$10.0 &&0.4862 & 0.4864 & 0.4773 & 0.4853 &12.3$\pm$5.7 & 12.7$\pm$6.0 & 11.5$\pm$13.9\\
$V_{\rm disp,peak}$ && 0.4897 & 0.3894 & 0.4889 & 0.3887 & 0.4874 & 0.3875 & 0.4887 & 0.3885 &17.1$\pm$4.8 & 18.5$\pm$6.6 & 18.9$\pm$11.4 &&0.5039 & 0.5033 & 0.4995 & 0.5028 &12.5$\pm$5.4 & 12.5$\pm$6.2 & 13.4$\pm$13.7\\
$V_{\rm disp,80\%}$ && 0.4766 & 0.3754 & 0.4771 & 0.3755 & 0.4698 & 0.3710 & 0.4759 & 0.3747 &14.4$\pm$4.8 & 15.7$\pm$6.6 & 16.1$\pm$8.5 &&0.4948 & 0.4958 & 0.4832 & 0.4942 &13.0$\pm$5.7 & 13.8$\pm$6.4 & 14.7$\pm$14.2\\
\cline{3-13}\cline{15-21}
$\phi$ && 0.4795 & 0.3803 & 0.4721 & 0.3744 & 0.4854 & 0.3854 & 0.4744 & 0.3763 &20.6$\pm$5.1 & 20.7$\pm$6.4 & 24.7$\pm$8.2 &&0.4931 & 0.4888 & 0.4887 & 0.4888 &15.5$\pm$6.0 & 15.8$\pm$6.3 & 17.0$\pm$13.5\\
$\psi_1$ && 0.4722 & 0.3721 & 0.4717 & 0.3713 & 0.4675 & 0.3696 & 0.4710 & 0.3710 &15.0$\pm$5.1 & 16.3$\pm$6.3 & 15.7$\pm$10.1 &&0.4865 & 0.4863 & 0.4779 & 0.4853 &12.6$\pm$5.8 & 12.9$\pm$6.2 & 13.1$\pm$13.6\\
$\psi_2$ && 0.4759 & 0.3750 & 0.4760 & 0.3748 & 0.4698 & 0.3711 & 0.4749 & 0.3741 &15.3$\pm$5.1 & 16.8$\pm$7.1 & 15.5$\pm$9.6 &&0.4894 & 0.4889 & 0.4823 & 0.4881 &14.0$\pm$5.6 & 14.4$\pm$6.4 & 15.2$\pm$13.1\\
$\psi_3$ && 0.4728 & 0.3725 & 0.4725 & 0.3720 & 0.4680 & 0.3700 & 0.4717 & 0.3716 &14.4$\pm$5.1 & 15.6$\pm$6.7 & 14.6$\pm$10.5 &&0.4874 & 0.4874 & 0.4788 & 0.4864 &12.5$\pm$5.8 & 13.1$\pm$6.0 & 13.5$\pm$13.7\\
$\psi_4$ && 0.4777 & 0.3763 & 0.4781 & 0.3764 & 0.4708 & 0.3718 & 0.4768 & 0.3756 &14.9$\pm$5.1 & 16.5$\pm$7.2 & 15.7$\pm$10.1 &&0.4963 & 0.4971 & 0.4848 & 0.4956 &13.3$\pm$5.4 & 13.9$\pm$6.1 & 15.6$\pm$13.5\\
$\psi_5$ && \cellcolor{blue!25}\textbf{0.4695} & \cellcolor{blue!25}\textbf{0.3700} & \cellcolor{blue!25}\textbf{0.4703} & \cellcolor{blue!25}\textbf{0.3703} & \cellcolor{blue!25}\textbf{0.4649} & \cellcolor{blue!25}\textbf{0.3676} & \cellcolor{blue!25}\textbf{0.4694} & \cellcolor{blue!25}\textbf{0.3699} &12.2$\pm$4.9 & 12.5$\pm$6.8 & \cellcolor{blue!25}\textbf{12.4}$\pm$10.6 &&\cellcolor{blue!25}\textbf{0.4831} & \cellcolor{blue!25}\textbf{0.4839} & \cellcolor{blue!25}\textbf{0.4748} & \cellcolor{blue!25}\textbf{0.4827} &11.2$\pm$6.0 & 11.1$\pm$6.5 & 13.2$\pm$13.3\\
$\psi_6$ && 0.4735 & 0.3731 & 0.4747 & 0.3737 & 0.4682 & 0.3699 & 0.4736 & 0.3731 &12.0$\pm$5.1 & 12.3$\pm$6.8 & 17.2$\pm$10.9 &&0.4889 & 0.4900 & 0.4803 & 0.4888 &11.5$\pm$5.5 & 11.2$\pm$6.2 & 13.4$\pm$12.6\\
\enddata
\end{deluxetable*}
\end{rotatetable*}

\end{document}